\documentclass[12pt]{article}
\usepackage[utf8]{inputenc}
\usepackage[english]{babel}
\usepackage{amsmath, amssymb, graphicx}

\usepackage{hyperref}
\usepackage{braket}
\usepackage{color}
\usepackage{xcolor}
\usepackage[normalem]{ulem}
\usepackage{braket}
\usepackage{multirow}
\usepackage{subcaption}

\newcommand{\be}{\begin{equation}}
\newcommand{\ee}{\end{equation}}
\newcommand{\bea}{\begin{eqnarray}}
\newcommand{\eea}{\end{eqnarray}}

\newcommand{\dd}{{\mathrm{d}}}

\newcommand{\fnm}{\footnotemark}
\newcommand{\fnt}{\footnotetext}
\newcommand{\RomanNumeralCaps}[1]
{\MakeUppercase{\romannumeral #1}}

\newcommand{\const}{\textrm{const}}

\graphicspath{{img}}

\begin{document}

	\begin{center}
		
		\large \bf
		Probing the holographic model of  $\mathcal{N} =4$ SYM rotating quark-gluon plasma.\\
	\end{center}

	\vspace{15pt}

	\begin{center}
		
		\normalsize\bf
		Anastasia Golubtsova\fnm[1]\fnt[1]{golubtsova@theor.jinr.ru}$^{, a,b}$
		and
		Nikita Tsegelnik\fnm[2]\fnt[2]{tsegelnik@theor.jinr.ru}$^{, a}$
		
		\vspace{7pt}
		
		\it (a) \ \ \ \ Bogoliubov Laboratory of Theoretical Physics, JINR,\\
		Joliot-Curie str. 6,  Dubna, 141980  Russia  \\
		
		(b) \ \ \ \ Steklov Mathematical Institute, Russian Academy of Sciences,\\ 
		Gubkina str. 8, Moscow,  119991 Russia\\
		
	\end{center}
	
	\vspace{15pt}

	\begin{abstract}
		We study Wilson loops in holographic duals of the $\mathcal{N}=4$ SYM quark-gluon plasma. For this we consider the Schwarzschild-$AdS_5$ and Kerr-$AdS_5$ black holes, 
		which are dual to the  non-rotating and  rotating  QGPs, correspondingly.  From temporal Wilson loops we find the heavy quark potentials in both backgrounds. For the temperature above the critical one we observe the Coulomb-like behaviour of the potentials. We find that increasing the rotation the interquark distance decreases, we also see that the increase of the temperature yields the similar behaviour. Moreover, at high temperatures values of the potentials in Kerr-$AdS_5$ are close to that one calculated in the  Schwarzschild-$AdS_5$ black hole. We also explore holographic light-like Wilson loops from which the jet-quenching parameters of a fast parton propagating in the QGP are extracted. We find that the rotation increases the value of the jet-quenching parameter. However, at high temperatures the jet-quenching parameters have a cubic dependence on the temperature as for the AdS black brane.
	\end{abstract}

	\tableofcontents
	\newpage
	\section{Introduction}
	
	Recently, much interest has been paid to study and understand a rotating quark-gluon plasma (QGP). It can be created in non-central heavy-ion collisions, where large initial orbital momentum of ions is partially transferred to the created medium, that leads to the relativistic rotation~\cite{Becattini:2008,QGPR2}. The space-time structure of the vorticity field, which also arises in non-central heavy-ion collisions,  may have non-trivial geometrical features, like femto-vortex sheets~\cite{Baznat:2016} or elliptic vortex rings~\cite{Ivanov:2020,Tsegelnik:2022}. The non-zero vorticity may result in different effects, for instance, the chiral vortical effect (for a review, see ~\cite{Kharzeev:2015znc}). Unfortunately, there is no direct way to investigate the QGP, so different probes are used to extract the information about the plasma properties.
	
	One of these probes is the global polarization of $\Lambda$-hyperons. Being produced in a rotating medium, particles with spin obtain a polarization that depends on the magnitude of rotation~\cite{Becattini:2013, Becattini:2020}. In fact, by virtue of the $\mathcal{P}$-violation in the weak decay $\Lambda \rightarrow {\rm p} + \pi$, the angular distribution of the detected protons depends on the orientation of the $\Lambda$'s spin. In other words, measuring the proton distributions and restoring the polarization of the $\Lambda$-hyperons, it is possible to estimate the magnitude of the QGP rotation. This experiment was carried out by the STAR collaboration~\cite{Abelev:2007zk,STAR:2017ckg}. Surprisingly, the extracted averaged vorticity value is $\omega \approx 10^{22}\,$s$^{-1}$, which leads to the hypothesis that QGP is the fastest rotating fluid ever observed in nature~\cite{STAR:2017ckg, Petersen:2017}.
	
	In a series of experiments~\cite{PHENIX:2004vcz, Arsene:2005, Back:2005, Adams:2005} it was found that hadron spectra with high transverse momenta $p_{\rm T}$ are suppressed in the medium. The suppression of elliptical flows $v_2$ was also observed. This may indicate that the medium formed in heavy-ion collisions is dense and non-transparent. The increase of the nuclear modification factor $R_{\rm AA}$, which observed at experiments, also predicts that the QGP is an opaque fluid. 

	Since the experiments also indicate that the quark-gluon plasma produced in HIC is a strongly-coupled fluid \cite{PHENIX:2004vcz}, it's quite reasonable to examine this system in the framework of the holographic duality \cite{CSLMRW,AREF,DGRT}. 
	In this approach the object of study is replaced by $\mathcal{N}=4$ SYM plasma, that is much more simpler and provides a qualitative insights of the strongly coupled regime. 
	It worth to be noted that at finite temperature strongly coupled $\mathcal{N} = 4$ SYM and QCD above the deconfinement temperature have much common.
	At high temperature lattice simulations show that the stress tensor becomes traceless, that may indicate on a conformal symmetry \cite{Bazavov:2014}.
	
	Note that $\mathcal{N} =4$ SYM defined on $R\times \mathbb{R}^3$  at zero temperature doesn't have a confinement-deconfinement phase transition. The holographic calculations in backgrounds with flat boundaries also predict that there is no confinement-deconfinement phase transition in the dual theory on $R\times \mathbb{R}^3$\cite{Rey:1998ik,Maldacena:1998im,Rey:1998bq,Brandhuber:1998bs, Brandhuber:1999jr}.
	However, the situation changes if one discusses  $\mathcal{N} =4$ SYM on $R\times \mathbb{S}^3$. In \cite{Sundborg:1999ue}  it was shown that  a first order phase transition occurs in  
	the free $\mathcal{N} =4$  SYM on $R\times \mathbb{S}^3$ at the Hagedorn temperature,  in \cite{Spradlin:2004pp} it was discussed for the one-loop order in the weak coupling expansion. 
	Moreover, using the integrability the Hagedorn temperature was calculated at any value of the ’t Hooft coupling in \cite{Harmark:2017yrv}.

	From the holographic point of view the strongly coupled $\mathcal{N} =4$  SYM on $R\times \mathbb{S}^3$ at finite temperature is dual
	to a 5d AdS black hole with a conformal boundary $R\times \mathbb{S}^{3}$, i.e. a spherical horizon. In turn, a 5d AdS black hole with a conformal boundary $R\times \mathbb{S}^{3}$ has the first order Hawking-Page phase transition, which according to the holographic dictionary corresponds to the deconfinement phase transition in the dual theory  \cite{Witten,Marolf:2013ioa}. Thus, the quark-gluon plasma state at equilibrium can be associated to the AdS black hole with a larger radius.
	
	Following the holographic dictionary, a rotating AdS black hole with a spherical horizon is a gravitational dual to the rotating $\mathcal{N}=4$ SYM plasma \cite{BLLM,BLMM,McInnes:2018hid,NataAtmaja:2010hd}. 
	Like  Schwarzschild-AdS black holes, rotating AdS black holes also have a Hawking-Page phase transition  \cite{HHT,HReal,Gibbons:2004ai}, which corresponds to a phase transition in the dual theory. Note that the phase transition in  Kerr-$AdS_{5}$  happens for certain values of the rotational parameters.  If at least one of the rotational parameters is large enough then the phase transition disappears \cite{Arefeva:2020jvo}. The calculations of the critical temperature $T_{c}$ in the Kerr-$AdS_{5}$ background predict that  $T_{c}$ decreases with the rotation \cite{Arefeva:2020jvo}. This is also observed in other holographic backgrounds for studies of the rotating quark-gluon plasma \cite{2010.14478, Braga:2022yfe}  and effective models \cite{Jiang:2016, Chernodub:2021, Fujimoto:2021}. However, it was shown in lattice calculations \cite{Braguta:2020,Braguta:2021} that rotating gluons increase the critical temperature, while the rotating fermions decrease it.

	In work  \cite{Bantilan:2018vjv} it was discussed holographic off-center heavy-ion collisions using the 5d Kerr-AdS black hole with two non-zero rotational parameters. 
	In \cite{Kaminski}  the authors extracted analytic expressions for transport coefficients (the shear viscosity, the longitudinal momentum diffusion coefficient,  etc.) and  calculated  quasi-normal modes for spinning black holes. Scalar perturbations of the Kerr-$AdS_{5}$ background with generic rotational parameters were also calculated in \cite{BarraganAmado:2021uyw}, where it was shown that  quasi-normal modes in  Kerr-$AdS_{5}$ at low temperature  can be encoded by zeros of the Painleve \RomanNumeralCaps{5} tau function. In \cite{Geytota:2021ycx} it was found circular pulsating string solutions in the 5d Kerr-AdS black hole with equal rotational parameters. 
		
	Recently, within the framework of the holographic duality the energy loss of heavy quarks were explored in the rotating quark-gluon plasma in \cite{Arefeva:2020jvo, Arefeva:2020knc, Golubtsova:2021agl}. 
	In these works  a holographic description of a rotating QGP is given by a 5d Kerr-AdS black hole, while the heavy quarks at finite temperature are associated by endpoints of open strings in the AdS black hole. 
	The endpoints are located on the conformal boundary of the black hole background, so the string hanging down to the black hole horizon. 
In \cite{2010.14478,Braga:2022yfe,Chen:2022obe} it was studied thermodynamic quantities, Polyakov and Wilson loops in a holographic rotating background, which mimic results for lattice simulations for a pure gluon rotating system and  a rotating system with $N=2$ flavors.
	
	In this paper we probe $\mathcal{N}=4$ SYM quark-gluon plasma on $R\times \mathbb{S}^{3}$ by Wilson loops using holography. 
	The dual description of the expectation value of the rectangular Wilson loop can be done in terms of the minimized Nambu-Goto action of a classical string, which both endpoints attached to the conformal boundary of the AdS black hole, while the string stretched down to the horizon \cite{Rey:1998bq, Brandhuber:1998bs}. In this work we focus on temporal and light-like Wilson loops. From the expectation value of a temporal Wilson loop  we extract a heavy quark-antiquark potential and explore the affect of the rotation on it.
		
	 The light-like Wilson loops can be used to study the jet-quenching phenomenon in the quark-gluon plasma, which is of interest since high-energy particles propagating through the QGP are strongly decelerated \cite{GYULASSY1990432, Baier:2000mf}.
	Following  \cite{LRW,Liu:2006he} the so-called  jet-quenching parameter $\hat{q}$ is defined as a coefficient of the $L^2$ term in the logarithm of a long light-like Wilson loop of width $L$.
	It encodes the description of  energy losses  for relativistic partons moving in the quark-gluon plasma. More precisely, the parameter $\hat{q}$ gives the squared average transverse 
	momentum exchange between the medium and highly energetic parton per unit path length.  In \cite{Giataganas:2012zy} the holographic calculations for the jet-quenching parameter was  generalized for the case of an arbitrary diagonal metric. Using holographic models in \cite{Rajagopal:2016uip, Brewer:2018mpk} it was analyzed a modification of an ensemble of jets, which propagate  through a strongly coupled plasma. Thus, using the AdS/CFT correspondence we are also able to find and analyze the jet-quenching parameter.
	The holographic computations in the planar AdS black brane background yield the following relation \cite{LRW, Liu:2006he}
	\be\label{JQintro}
	\hat{q} 
	= \frac{\pi^{3/2} \Gamma(\frac{3}{4})}{\Gamma(\frac{5}{4})} \sqrt{\lambda} T^3.
	\ee
		
	It worth to be noted that the jet-quenching parameter $\hat{q}$ in \eqref{JQintro} is not proportional to the "number of scattering centers", which is $\propto N^{2}_{c}$. Moreover,  the value of $\hat{q}$ in QCD is smaller than that one predicted by \eqref{JQintro}. 	
	
	In this work we show that the jet-quenching parameters in the Schwarzschild-$AdS_{5}$ and Kerr-$AdS_{5}$ at high temperature have the same dependence on $T$ as for the AdS black brane \eqref{JQintro}. 
We also find that the rotation increases the values of $\hat{q}$.

	The paper is organized as follows. In Sec.2 we start with a review of the Schwarzschild-$AdS_5$ and Kerr-$AdS_5$ black hole solutions. Then we briefly discuss the calculation of rectangular Wilson loops in holography. In Sec.3  we calculate the expectation values of temporal Wilson loops in the Schwarzschild-$AdS_5$ and Kerr-$AdS_5$ black hole backgrounds and then analyze the corresponding quark-antiquark potentials. In Sec.4 we evaluate light-like Wilson loops and estimate the jet-quenching parameters in the Schwarzschild-$AdS_5$ and Kerr-$AdS_5$ black holes. In Sec.5 we conclude and give a discussion.

	\setcounter{equation}{0}
	\section{Setup}
	\subsection{Gravity backgrounds}
	We consider a  5-dimensional gravity theory with a negative cosmological constant $\Lambda$
	\be\label{5daction}
	S = \frac{1}{2\kappa^{2}}\int d^{5}x\sqrt{-g}\left(R_{5} -2\Lambda \right).
	\ee
	The Einstein equations following from (\ref{5daction}) are
	\be\label{EinEq}
	R_{\mu\nu} = \frac{\Lambda}{3} g_{\mu\nu},
	\ee
	where we suppose  $\Lambda=-6/\ell^{2}$. The simplest black hole solution of a mass $M$ and a spherical horizon to eqs. (\ref{EinEq}) is the  Schwarzschild-$AdS_5$ black hole with the metric
	\begin{align}\label{Schw.1psi}
	ds^{2} = - \frac{f(r)}{r^{2}}\dd t^{2}  +\frac{ r^{2} }{f(r)}\dd r^{2} +  r^{2} \left(  \dd\theta^{2} + \sin^{2}\theta\dd\phi^{2} + \cos^{2}\theta \dd\psi^{2}\right),
	\end{align}
	where the function $f(r)$ is
	\be\label{deltaSchwpsi}
	f(r) = r^{2} + \ell^{-2} r^{4} -2M.
	\ee
	In (\ref{Schw.1psi}) the angular coordinates are defined as $ 0\leq \theta \leq \pi/2$, $0\leq \phi, \psi \leq 2\pi$.
	It worth to be noted, that the horizon of the black hole is defined as a greater root of the equation $f(r)/r^2 =0$, thus we have
	\be
	r_{h}=  \frac{ \ell \sqrt{\sqrt{8 \ell^{-2} M+1}-1}}{\sqrt{2}}.
	\ee
	The Hawking temperature of the black hole \eqref{Schw.1psi} is given by
	\be\label{eq:setup:Th-Schwarz}
	T_{\rm H} = \frac{2r_{h}^2 + \ell^2}{2\pi r_{h}\ell^2}.
	\ee
	
	Another black hole solution with a spherical horizon to eqs. \eqref{EinEq} is the Kerr-$AdS_{5}$ black hole with arbitrary rotational parameters $a$ and $b$ (in the static-at-infinity frame \cite{Gibbons:2004ai})
	\begin{align}\label{eq:metric:asymptAdSKerr-ab} 
	ds^{2} \simeq &-\left(1 + y^2 \ell^{-2}\right) \dd T^2 + \frac{\dd y^2}{1 + y^2 \ell^{-2} - \frac{2M}{\Delta^2 y^2}} + \frac{2M}{\Delta^3 y^2} \left(\dd T - a \sin^2 \Theta \dd \Phi - b \cos^2 \Theta \dd \Psi \right)^2 \nonumber\\
	&+ y^2 \left(\dd \Theta^2 + \sin^2 \Theta \dd \Phi^2 + \cos^2 \Theta \dd \Psi^2\right),
	\end{align}
	with
	\begin{equation}\label{eq:delta:asymptAdSKerr-ab}
	\Delta = 1 - a^2 \ell^{-2} \sin^2 \Theta - b^2 \ell^{-2} \cos^2 \Theta.
	\end{equation}
	In (\ref{eq:metric:asymptAdSKerr-ab})-(\ref{eq:delta:asymptAdSKerr-ab}) the angular coordinates run as for the Schwarzschild-$AdS_5$: $0\leq\Theta \leq \frac{\pi}{2}$, $0\leq\Phi, \Psi \leq 2\pi$. 
	The horizon $y_{+}$ of the Kerr-$AdS_5$  black hole is a greater root of the equation
	\be
	1 + y^2 \ell^{-2} - \frac{2M}{\Delta^3 y^2}= 0.
	\ee
	Correspondingly, the Hawking temperature reads
	\be\label{eq:setup:Th}
	T_{\rm H}= \frac{1}{2\pi}\left(y_{+}(1+ y^{2}_{+}\ell^{-2})\Bigl(\frac{1}{y^{2}_{+} + a^{2}} + \frac{1}{y^{2}_{+} + b^{2}}\Bigr) - \frac{1}{y_{+}}\right).
	\ee
	It's easy to see, that for $a=b=0$ the Kerr-$AdS_5$ metric~\eqref{eq:metric:asymptAdSKerr-ab} comes to the Schwarzschild-$AdS_5$~\eqref{Schw.1psi} background, i.e. $y_+\big\vert_{a=b=0} = r_h$. 
	
\begin{figure}[!htb]
		\centering
		\includegraphics[width=0.5\linewidth]{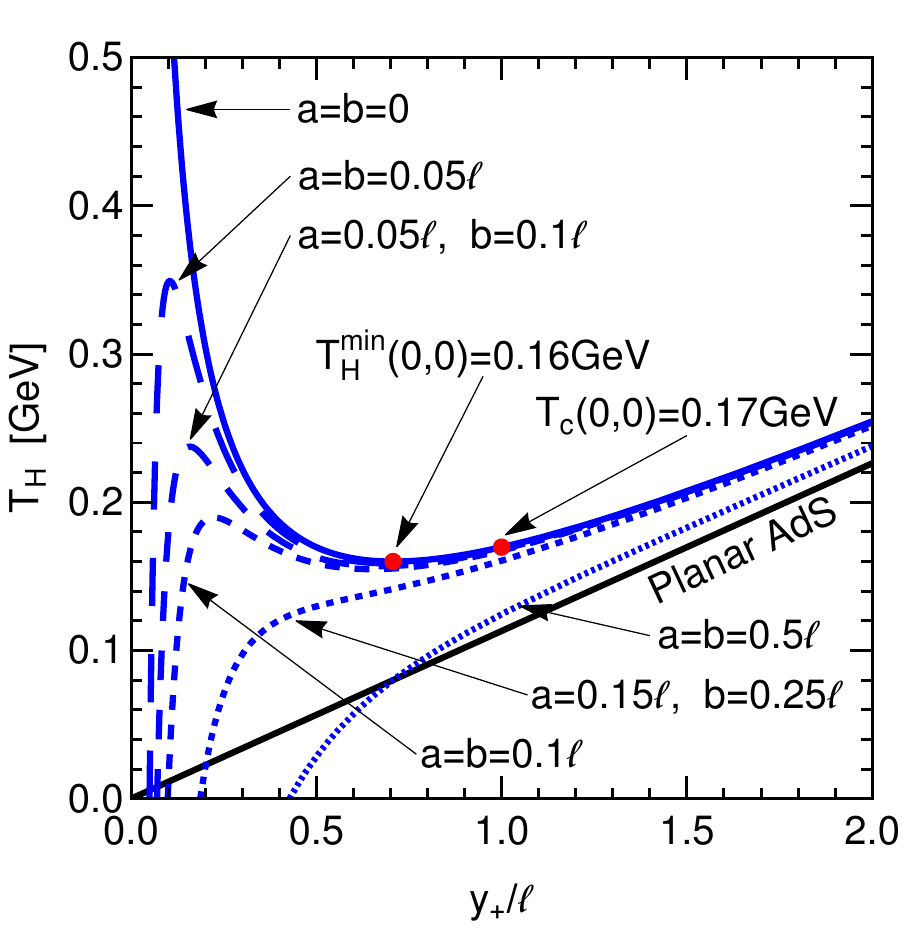}
		\caption{The Hawking temperature $T_{\rm H}$  as a function of $y_{+}/\ell$ ($r_{h}/\ell$). The case of the Schwarzschild-$AdS_5$ black hole ($a=b=0$) is shown by a blue solid curve,  the  Hawking temperatures for the Kerr-$AdS_{5}$ background for various values of the rotational parameters are shown by dashed  curves from top to bottom according to increasing values of $a$ and $b$. The Hawking temperature for the AdS black hole with planar horizon is depicted by the black solid line.}
		\label{fig:Th-rh}
	\end{figure}

	The dependence of the Hawking temperature  on  $y_{+}/\ell$ ($r_{h}/\ell$) is shown in Fig.~\ref{fig:Th-rh}. The rotational parameters belong to the range $0 \leq a, b \leq \ell$, so it is useful  to take a fraction of $\ell$ as a value of $a$ or $b$.
	We see that  both the Schwarzschild-$AdS_5$ and Kerr-$AdS_5$ black holes have minima of the Hawking temperature $T^{\rm min}_{\rm H}$. In the case of the Schwarzschild-$AdS_5$ black hole $T^{\rm min}_{\rm H}$ is defined by 
	\be
	T_{\rm H}^{\rm min} = \frac{\sqrt{2}}{\pi \ell}, \quad \textrm{with}\quad r_{h} = \frac{\ell}{\sqrt{2}}.
	\ee
	Note that the black hole solution doesn't exist for $T< T^{\rm min}_{\rm H}$. 
	In our calculations we set $\ell=0.55\,$fm, so $T^{\rm min}_{\rm H} \approx0.16\,$GeV. Above this point, there are two possible values of the temperature, corresponding to the small ($r_h < \ell/\sqrt{2}$) and big ($r_h > \ell/\sqrt{2}$) black holes, but only the latter is allowed as a stable equilibrium \cite{Marolf:2013ioa}. The Hawking-Page phase transition occurs at the temperature $T_c \geq 3/(2\pi\ell)\approx0.17\,$GeV.  Following the holographic dictionary  the Hawking-Page phase transition corresponds to the confinement-deconfinement phase transition \cite{Witten}.
	
	For the Kerr-$AdS_{5}$ black hole the Hawking-Page phase transition also takes place. However, the presence of rotation changes the behaviour of the temperature: below some critical values of the rotational  parameters the  temperature is three-valued function on $y_+$. From the other hand, a stronger rotation leads to the absence of the temperature ambiguity, and, hence, the Hawking-Page phase transition disappears~\cite{Arefeva:2020jvo}.
	
	In Fig.~\ref{fig:Th-rh} we also compare  the Hawking temperatures of the 5d AdS black holes with planar and spherical horizons. We see that for the same values of the horizons, $T_{\rm H}$ of the black hole with spherical symmetry has a greater value than that one for the planar $AdS_5$ black hole. The spherical AdS black hole with a large horizon, corresponding a high temperature, behaves  similar to the planar AdS black holes. 
	
	Note that the conformal boundary for both solutions~\eqref{Schw.1psi}, \eqref{eq:metric:asymptAdSKerr-ab} is defined at infinity of the holographic coordinates $r\to+\infty$ ($y\to +\infty$) and has the form\footnote{For the Kerr-$AdS_5$ the coordinates in~\eqref{BNDR} should be capital.}
	\be\label{BNDR}
	ds^2 = - \dd t^2 + \dd\theta^2 +\sin^2\theta \dd\phi^2 +\cos^2\theta\dd\psi^2.
	\ee
	
	\subsection{Wilson loops in holography}
	
	Following the holographic prescription the expectation value of the Wilson loop on the contour $\mathcal{C}$ can be calculated using a Nambu-Goto action of an open string in a holographic background \cite{Rey:1998ik,Maldacena:1998im}
	\be\label{WLholo}
	\Braket{W(\mathcal{C})}=e^{-S_{\rm NG}},
	\ee
	where $S_{\rm NG}$ is a regularized action of the string.
	Hence, consider a string  which is governed by the Nambu-Goto action
	\be\label{defNGact}
	S_{\rm NG}= \frac{1}{2\pi \alpha'}\int \dd \sigma \dd \tau \sqrt{-\det(g_{\alpha\beta})},
	\ee
	where $\sigma$ and  $\tau$ parametrized the string worldsheet, $(g_{\alpha\beta})$ is the induced metric on the  worldsheet
	\be\label{defindm}
	g_{\alpha\beta} = G_{MN} \partial_{\alpha}X^{M}\partial_{\beta}X^{N},
	\ee
	$G_{MN}$ is a spacetime metric,  $X^{M}$ are embedding coordinates,  $\alpha,\beta$ are worldsheet indices. 
	To consider a temporal recangular Wilson loop, one should take one temporal and one spacial coordinate to parametrize the string worldsheet.
	
	It's known that the interquark potential is related to the expectation value of the static temporal 
	Wilson loop as follows
	\be
	\Braket{W(\mathcal{C})}\sim e^{-\mathcal{T}V(L)},
	\ee
	where  the distance between quarks  $L$ and the temporal extent of the Wilson loop $\mathcal{T}\to \infty$.
	Thus, taking into account (\ref{WLholo}) the quark-antiquark potential can be found in the following way
	\be\label{VqqSng}
	V_{q\bar{q}} = \frac{S_{\rm NG}}{\mathcal{T}}|_{\mathcal{T}\to \infty}.
	\ee
A generalization to the finite temperature case was suggested in \cite{Rey:1998bq,Brandhuber:1998bs}. In  the work \cite{Brandhuber:1999jr}  the quark-antiquark potential was explored in the rotating D3-brane background.
	
	Note that the Cornell potential \cite{Eichten:1976,Eichten:1978}  includes the Coulomb term, which dominates at short distances, and the linear confining term
	\begin{equation}\label{eq:cornell}
	V_{q\bar{q}} =  \sigma L -\frac{\kappa}{L},
	\end{equation}
	where $L$ is the interquark distance, $\kappa$ and $\sigma$ are the Coulomb strength and string tension parameters, 
	correspondingly. 
	In the confined phase the expectation value of the Wilson loop reproduces an area law
	\be
	\Braket{W(\mathcal{C})}\sim e^{-\sigma L T}=e^{-\sigma {\rm Area}(\mathcal{C})}.
	\ee
	
	Using the expectation value of the light-like Wilson loop on the contour $\mathcal{C}$ in the adjoint  representation one is able to find
	the jet-quenching parameter $\hat{q}$ for a fast parton \cite{LRW,Liu:2006he}
	\be\label{eq:JQmain}
	\Braket{ W^{A}(\mathcal{C} )} \approx \exp\Bigl[- \frac{1}{4\sqrt{2}} \hat{q} L^{-} L^2\Bigr],
	\ee
	where $L^{-}$ is a large side of the rectangular contour $\mathcal{C}$ and $L$ is a short side.
	At the same time, the Wilson loop operator in the adjoint representation is related to the Wilson loop operator in the fundamental representation as follows
	\be
	\Braket{W^{A}(\mathcal{C})} \approx \Braket{W^F(\mathcal{C})}^2.
	\ee
	Following the holographic dictionary  (\ref{WLholo}), we have
	\be\label{eq:JQmain2}
	\Braket{W^{F}(\mathcal{C})} = e^{-S_{\rm NG}}.
	\ee
	Taking into account (\ref{eq:JQmain}) we find the  relation for the jet-quenching parameter
	\be\label{eq:JQmain3}
	\hat{q} = \frac{4 \sqrt{2}}{L^{-} L^2} S_{\rm NG}.
	\ee

	\setcounter{equation}{0}
	\section{Holographic Wilson loops}
	
	\subsection{Wilson loop in Schwarzschild-$AdS_5$ black hole}
	It is instructive to start with a non-rotating case of the holographic background, so, first, we consider
	a holographic  Wilson loop in  the 5d Schwarzschild-AdS black hole with a spherical horizon (\ref{Schw.1psi})-(\ref{deltaSchwpsi}).
	
\begin{figure}[!htb]
		\centering
		\includegraphics[width=0.49\linewidth]{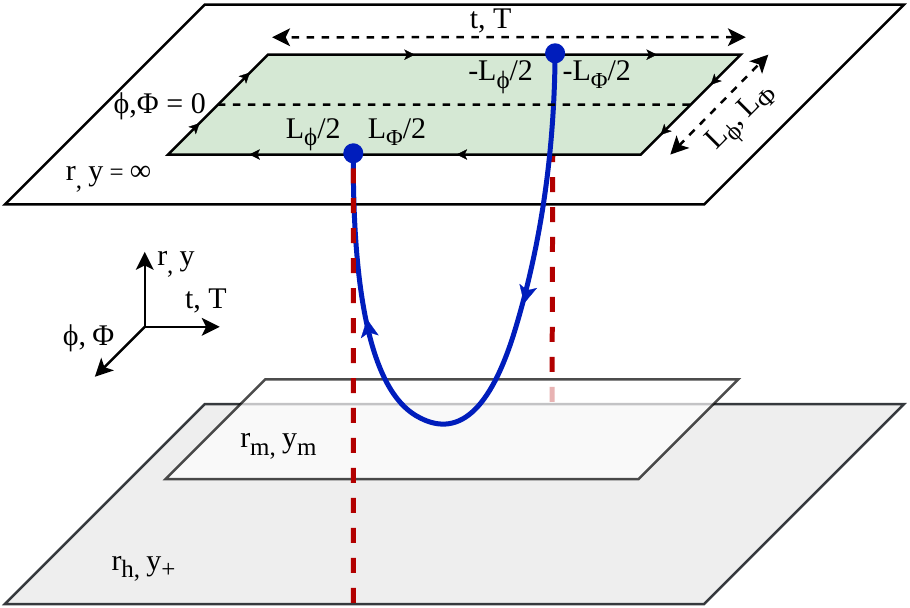}
		\caption{The schematic illustration of the holographic Wilson loop configuration. The string endpoints are located at $\phi=\pm \frac{L_{\phi}}{2}$. The red dashed lines depict the configuration of the free quarks.}
		\label{fig:string}
	\end{figure}

	Parametrizing the worldsheet of the static string  in the following way:
	\begin{equation}\label{eq:Schwarz:parametrization}
	\tau = t, \qquad \sigma =  \phi, \qquad \phi \in [0, 2\pi L_\Phi],\quad r=r(\phi),
	\end{equation}
	we get non-zero components of the induced metric  (\ref{defindm})
	\be\label{eq:Schwarz:metric:induced}
	g_{\tau\tau} = G_{tt} = -\frac{f(r)}{r^2},\quad
	g_{\sigma\sigma} = G_{\phi\phi} + r'^2 G_{rr} = r^2 \left(\sin^2 \theta+ \frac{r'^2}{f(r)}\right),
	\ee
	where $f(r)$ is defined by (\ref{deltaSchwpsi})  and we denoted $r' \equiv \dd r/\dd\phi$.
	The boundary conditions for the string endpoints are given by
	\be\label{bndr:cond}
	r\left(\phi = -\frac{L_{\phi}}{2}\right)= r\left(\phi = \frac{L_{\phi}}{2}\right) = \infty.
	\ee
	In Fig.~\ref{fig:string} we show the string configuration  for (\ref{eq:Schwarz:parametrization}),(\ref{bndr:cond}).
	
	Eqs.(\ref{eq:Schwarz:parametrization})-(\ref{bndr:cond}) yield the following expression for the Nambu-Goto action (\ref{defNGact}) of the string in the Schwarzschild-$AdS_5$ background
	\begin{equation}\label{eq:Schwarz:wloop:action}
	S =   \frac{T}{2 \pi \alpha'} \int^{\frac{L_{\phi}}{2}}_{-\frac{L_{\phi}}{2}}d \phi\, \sqrt{f(r) \sin^2 \theta + r'^2}.
	\end{equation}
	From \eqref{eq:Schwarz:wloop:action} it is easy to find the  integral of motion
	\begin{equation}\label{eq:Schwarz:wloop:hamiltonian}
	\mathcal{H} = - \frac{\sin^2 \theta \sqrt{f(r)}}{ \sqrt{\sin^2 \theta + \frac{r'^2}{f(r)}}}= - \frac{\ell}{C}.
	\end{equation}
	The string has a turning point, which is defined by $r'|_{\phi_{m}}=0$, thus from (\ref{eq:Schwarz:wloop:hamiltonian}) we have 
	\be
	- \frac{\sin^2 \theta \sqrt{f(r)}}{ \sqrt{\sin^2 \theta + \frac{r'^2}{f(r)}}}=- \frac{\ell}{C},
	\ee
	so the  constant of integration is defined by 
	\be\label{defCconstInt}
	C = \frac{\ell}{ \sin\theta\sqrt{f(r)}} \big\vert_{r=r_m}
	\ee
	with $r_{m} = r(\phi_{m})$. Note that $r_{m}$ is located  above the horizon $r_{h}$, see Fig.~\ref{fig:string}.
	
	From eq.(\ref{eq:Schwarz:wloop:hamiltonian}) we find the equations of motion  represented as
	\begin{equation}\label{eq:Schwarz:wloop:eom}
	r'^2 = \sin^2 \theta f(r) \left( \frac{C^2 \sin^2 \theta f(r)}{ \ell^2} - 1 \right).
	\end{equation}
	Plugging (\ref{eq:Schwarz:wloop:eom}) into the Nambu-Goto action (\ref{eq:Schwarz:wloop:action}) and coming to the integration in terms of $r$, we obtain
	\begin{equation}\label{eq:Schwarz:wloop:action-2}
	S_{\rm NG} = \frac{T}{ \pi \alpha'} \int^{\infty}_{r_m} d r\, \frac{C \sin \theta \sqrt{f(r)}}{\sqrt{C^2 \sin^2 \theta f(r) - \ell^2}}.
	\end{equation}
	From the other hand,  we have the expression for the distance between quarks $L_\phi$ from (\ref{eq:Schwarz:wloop:eom}):
	\begin{equation}\label{eq:Schwarz:wloop:Lphi}
	\frac{L_\phi}{2} = \frac{ \ell}{\sin \theta} \int^{\infty}_{r_m} dr\, \frac{1}{\sqrt{f(r)} \sqrt{ C^2 \sin^2 \theta f(r) - \ell^2 }}.
	\end{equation}
	It worth to be noted that eq.\eqref{eq:Schwarz:wloop:action-2} has a divergence at the conformal boundary  $r\to +\infty$ of the spacetime \eqref{Schw.1psi} and we have to regularise (\ref{eq:Schwarz:wloop:action-2}). The renormalization procedure represents a subtraction of the "self-energy" of two free static quarks, which holographically correspond to the action of  the static straight strings stretched from the boundary $r=\infty$ to the horizon $r_h$:
	\begin{equation}\label{eq:Schwarz:wloop:free-action}	
	S_0 = \frac{T}{\pi \alpha'} \int_{r_h}^{\infty} dr\, \sqrt{-G_{tt} G_{rr}} 
	= \frac{T}{\pi \alpha'} \left(\int_{r_m}^{\infty} dr + r_m - r_h \right).
	\end{equation}
	Then taking into account (\ref{eq:Schwarz:wloop:free-action}) the regularized action takes the form
	\bea\label{eq:Schwarz:wloop:action-rn}
	S_{\rm NG}^{\rm ren} = S_{\rm NG} - S_0 = \frac{T}{ \pi \alpha'} \left(\int^{\infty}_{r_m} dr\,\left( \frac{C \sin \theta \sqrt{f(r)}}{\sqrt{C^2 \sin^2 \theta f(r) - \ell^2}} - 1\right) -r_m+r_h\right).
	\eea
	
	One can try to estimate the relation between $S^{\rm ren}_{\rm NG}$  \eqref{eq:Schwarz:wloop:action-rn}  and $L_{\phi}$ \eqref{eq:Schwarz:wloop:Lphi}. In order to find this we introduce the following notation in eqs. \eqref{eq:Schwarz:wloop:Lphi} and \eqref{eq:Schwarz:wloop:action-rn}
	\be
	S_{\rm NG}^{\rm ren} =\frac{T}{ \pi \alpha'} I_1(r_m,C),\quad L_{\phi} =2 I_2(r_m,C).
	\ee
	The derivatives of  these quantities with respect to $C$  are related in the following way
	\begin{equation}\label{eq:Schwarz:wloop:diffeq}
	\frac{\partial I_2(r_m,C)}{\partial C} = \frac{C}{\ell}  \frac{\partial I_1(r_m,C)}{\partial C}.
	\end{equation}
	Integrating LHS of eq. \eqref{eq:Schwarz:wloop:diffeq}, we obtain
	\be\label{eq:Schwarz:wloop:diffeq-int1}
	\int_0^{C} \frac{\partial I_2(r_m,C)}{\partial C} d C  = \frac{L_{\phi}}{2} + \frac{i }{\sin \theta} \int^{\infty}_{r_m} \frac{dr}{\sqrt{f(r)}},	
	\ee
	at the same time using integration by parts of RHS  \eqref{eq:Schwarz:wloop:diffeq} one has
	\be\label{eq:Schwarz:wloop:diffeq-int2}
	\int_0^{C} \frac{C}{\ell} \frac{\partial I_1(r_m,C)}{\partial C} d C = \frac{C}{\ell} \frac{ \pi \alpha'}{T}S^{\rm reg}_{\rm NG}- \frac{1}{\ell}\int_0^{C} I_1(r_m,C) d C,
	\ee
	where we define 
	\be\label{eq:newint}
	\int_0^{C} I_1(r_m,C) d C =
	\int_{r_m}^{\infty}dr\left(\left(\frac{\sqrt{C^{2}\sin^{2}\theta f(r) - \ell^2}}{\sin\theta\sqrt{f(r)}} - C\right)-\frac{i \ell}{\sin\theta\sqrt{f(r)}}\right) -C(r_m-r_h). 
	\ee
	Taking into account  \eqref{eq:Schwarz:wloop:diffeq} -(\ref{eq:newint})  we get the  following relation between the quantities $S_{\rm NG}$ and $L_{\phi}$
	\be	\label{eq:wl:schwarz:vqq-0}
	S^{\rm reg}_{\rm NG} = \frac{T}{\pi \alpha'}\frac{\ell}{C}\left(\frac{L_{\phi}}{2} + I_3(r_m,C)\right),
	\ee
	where
	\begin{equation}\label{eq:wl:schwarz:i3}
	\begin{aligned}
	I_3(r_m,C) &= \int_{r_m}^{\infty}dr\left(\frac{\sqrt{C^{2}\sin^{2}\theta f(r) - \ell^2}}{\ell \sin\theta\sqrt{f(r)}} - \frac{C}{\ell}\right) - \frac{C}{\ell}(r_m-r_h),\\
	\end{aligned}
	\end{equation}
	and $C$ is defined by (\ref{defCconstInt}).\\
	
	Plugging  (\ref{defCconstInt}), \eqref{eq:wl:schwarz:vqq-0}-\eqref{eq:wl:schwarz:i3} into \eqref{VqqSng} and doing some algebra, we find the following relation for the quark-antiquark potential
	\begin{equation}\label{eq:Schwar:Vqq}
	\begin{aligned}
	V_{q\bar{q}} &= \frac{\sqrt{\lambda}}{\pi \ell^2} \sin \theta \sqrt{f(r_m)} \left( \frac{L_{\phi}}{2} + I_3\right),\\
	I_3 &= \frac{1}{\sin \theta \sqrt{f(r_m)}} \left[\int_{r_m}^{\infty} \left(\sqrt{1 - \frac{f(r_m)}{f(r)}} - 1\right)dr - (r_m-r_h)\right],
	\end{aligned}
	\end{equation}
	where   $\sqrt{\lambda } = \ell^2/\alpha'$ and the distance between quarks $L_{\phi}$ is given by (\ref{eq:Schwarz:wloop:Lphi}) with $C$ defined in (\ref{defCconstInt}).
	
\begin{figure}[!htb]
		\centering
		\captionsetup[subfigure]{labelformat=empty}
		\begin{subfigure}{0.39\linewidth}
			\includegraphics[width=1\linewidth]{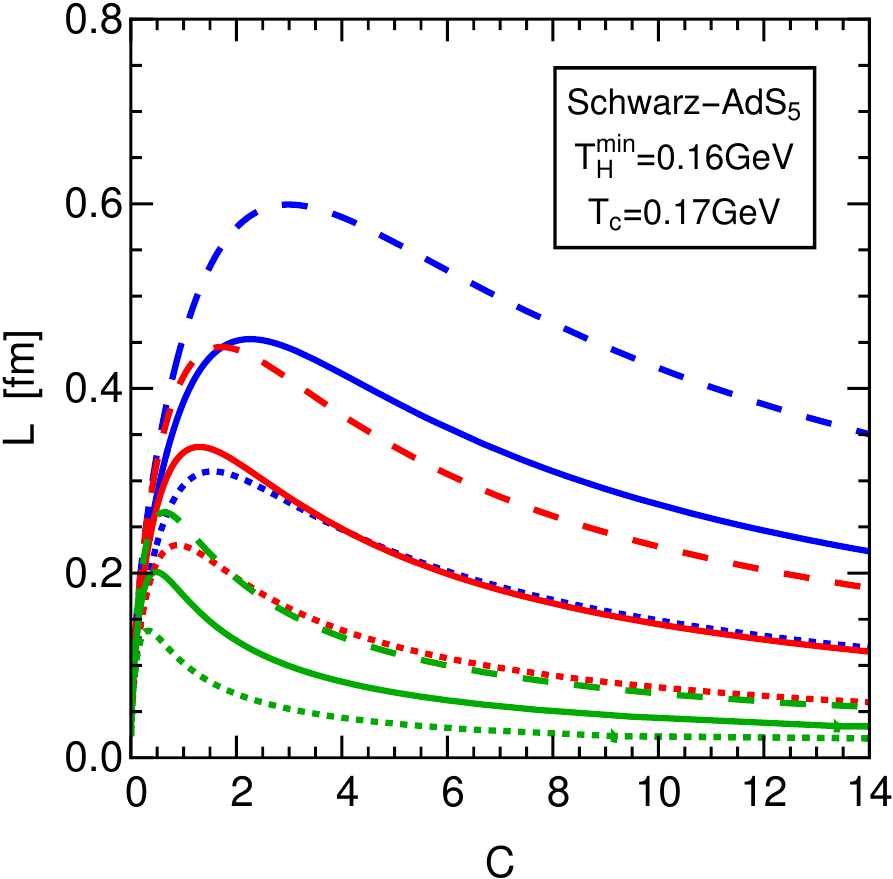}\vspace*{2pt}
			\caption{A}
		\end{subfigure}
		\begin{subfigure}{0.2\linewidth}
			\includegraphics[width=1\linewidth]{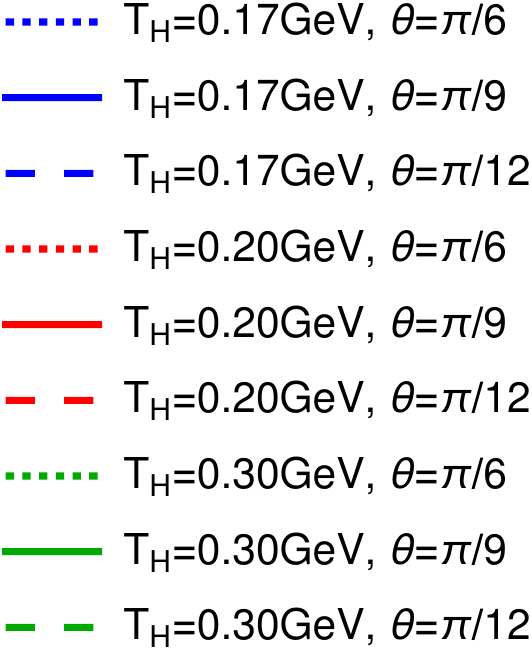}\vspace*{5em}
		\end{subfigure}
		\begin{subfigure}{0.39\linewidth}
			\includegraphics[width=1\linewidth]{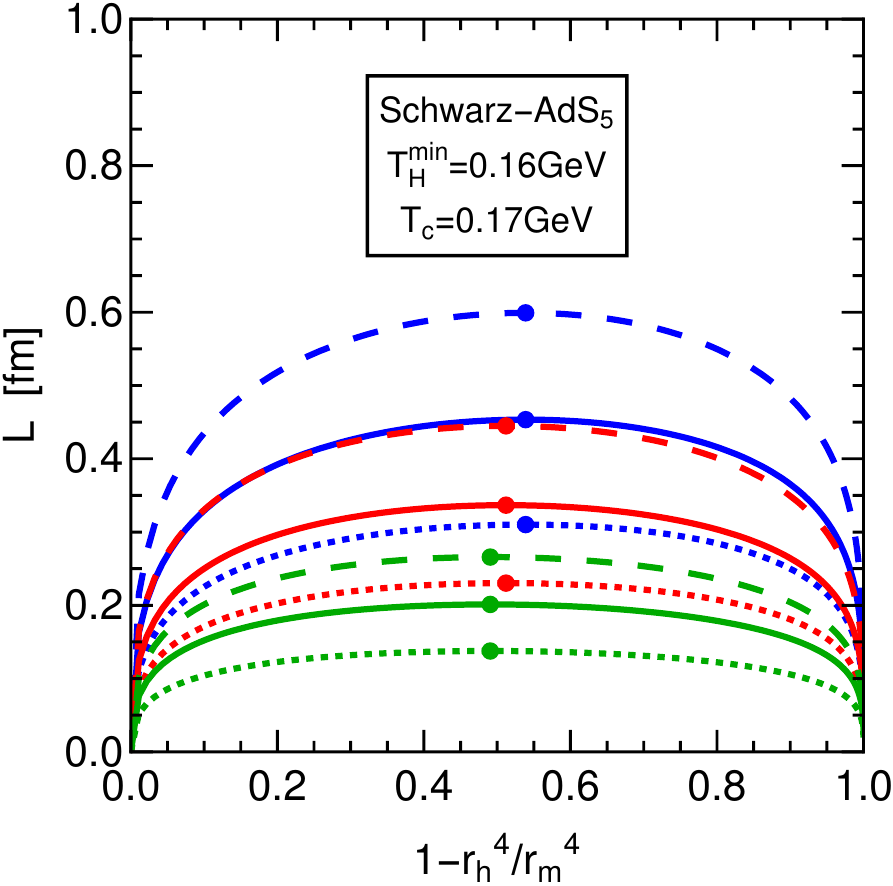}
			\caption{B}
		\end{subfigure}
		\caption{A) The distance  between quark-antiquark pair $L_{\phi}$ as a function of  the integration constant $C$~\eqref{defCconstInt}. B) The distance between quark-antiquark pair $L_{\phi}$ as a function of   $1 - \frac{r_h^4}{r^4_m}$.}
		\label{fig:LSchwar-C}
	\end{figure}

\begin{figure}[!htb]
		\centering
		\captionsetup[subfigure]{labelformat=empty}
		\begin{subfigure}{0.39\linewidth}
			\includegraphics[width=0.95\linewidth]{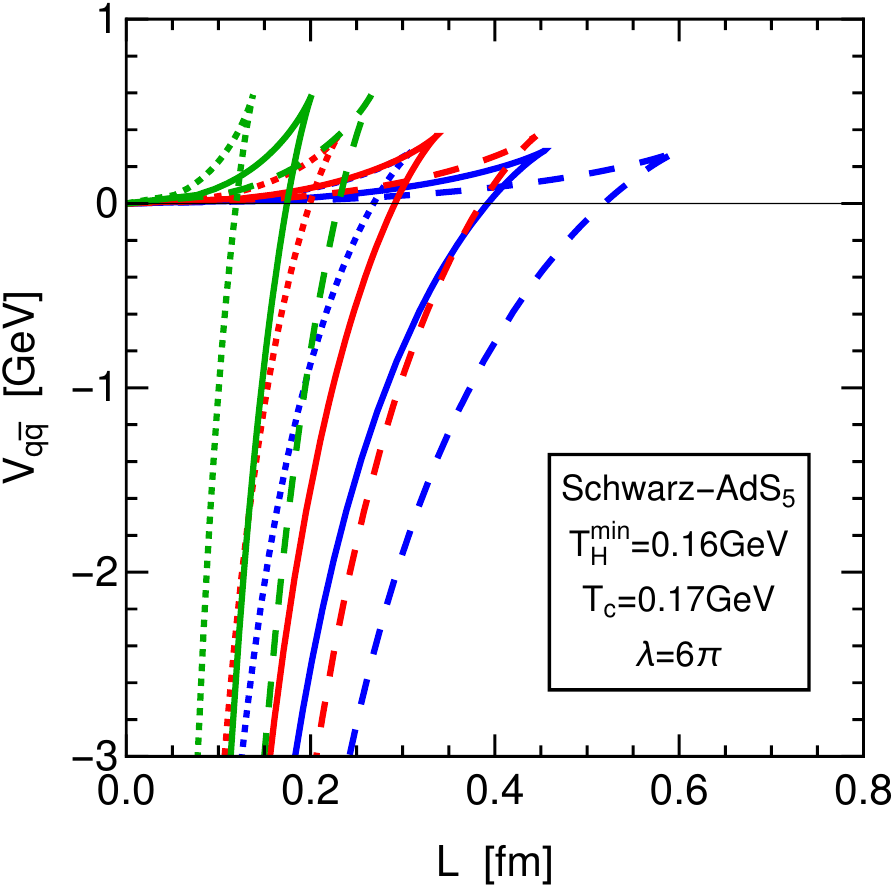}
			\caption{A}
		\end{subfigure}\hfill
		\begin{subfigure}{0.2\linewidth}
			\includegraphics[width=1\linewidth]{Vqq-L-Schwarz-leg}\vspace*{5em}
		\end{subfigure}\hfill
		\begin{subfigure}{0.39\linewidth}
			\includegraphics[width=1\linewidth]{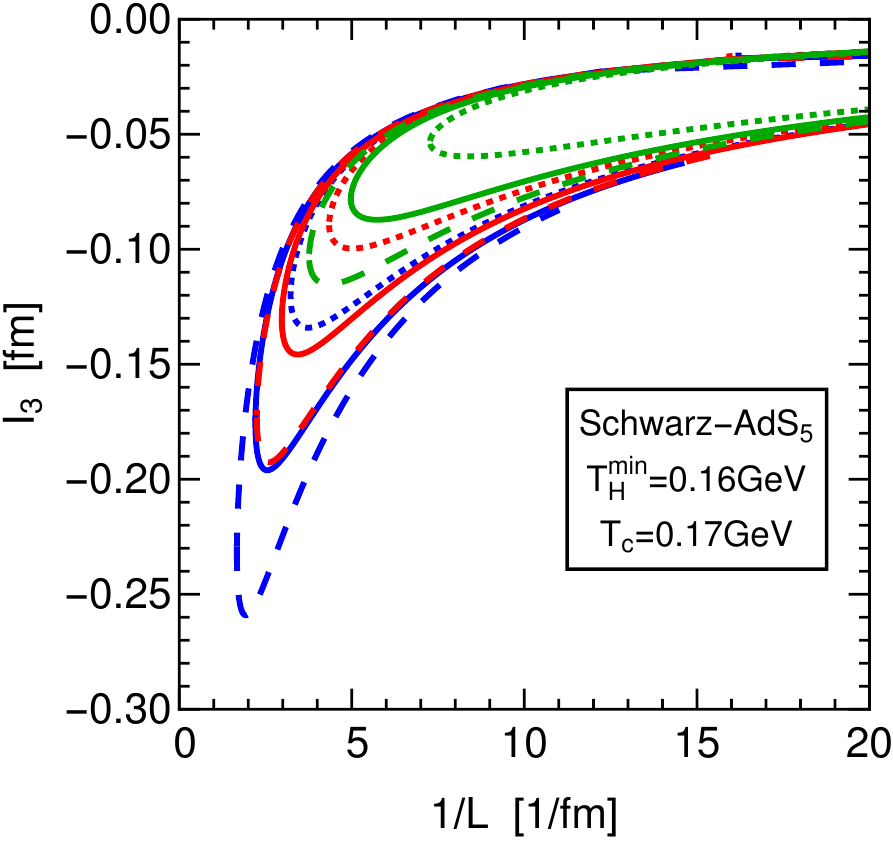}
			\caption{B}
		\end{subfigure}
		\caption{A) The quark-antiquark potential $V_{q\bar{q}}$ as a  function of the distance  $L_{\phi}$. B) The dependence of $I_3$ on the inverse distance $1/L_{\phi}$.}
		\label{fig:Vqq-Schwar}
	\end{figure}

	In Figs.~\ref{fig:LSchwar-C}-\ref{fig:Vqq-Schwar}  we present the numerical studies of the dependence of the quark-antiquark potential $V_{q\bar{q}}$ (\ref{eq:Schwarz:wloop:action-rn}) on the distance $L_{\phi}$ \eqref{eq:Schwarz:wloop:Lphi}.
	For all plots we perform numerical calculations at various temperatures keeping the ’t Hooft coupling fixed as $\lambda=6\pi$  and  varying the angle $\theta$. 
	In order to set the minimal Hawking temperature  $T_{\rm H}^{\rm min} = 0.16$ GeV, we put $\ell\approx0.55\,{\rm fm \,{\rm}}$. It worth to be mentioned, that the phase transition occurs at a slightly higher temperature, namely, at $T_{\rm c} = 3/(2 \pi \ell) \approx 0.17\,{\rm  GeV}$.
	
	The interquark distance  $L_{\phi}$  (\ref{eq:Schwarz:wloop:Lphi}) as a function of the integration constant $C$~\eqref{defCconstInt} is depicted in Fig.~\ref{fig:LSchwar-C}{\bf A}.  We observe that $L_{\phi}$ decreases as the temperature increases. One can also see that for a fixed temperature $T_{\rm H}$  the distance $L_{\phi}$ takes the smaller values while  $\theta$ increases.

	In Fig.~\ref{fig:LSchwar-C}{\bf B} we show the distance between the quark-antiquark pair $L_{\phi}$  (\ref{eq:Schwarz:wloop:Lphi}) as a function of the quantity  $1- \left(r_h/r_{m}\right)^4$. This plot illustrates the dependence of the interquark separation $L_{\phi}$ on the turning point $r_{m}$.
	As in the previous plot we observe that  the distance decreases with increasing temperature $T_{\rm H}$ and that the angle reduction leads to a decrease in the distance between quarks. 
	In Fig.~\ref{fig:LSchwar-C}{\bf B} we also see that $L_{\phi}$ increases until it reaches its maximal value $L_{\phi,\rm max}$ and then it decreases.	It is interesting that one is able to obtain the same value of $L_{\phi}$ tuning the temperature $T_{\rm H}$ and the angle $\theta$. 
	
	The behaviour of the quark-antiquark potential $V_{q\bar{q}}$  on the interquark distance $L_{\phi}$ is presented in Fig.~\ref{fig:Vqq-Schwar}{\bf A}. For the plot we choose $T_{\rm H}=0.17$ GeV, that corresponds to the deconfined phase. We see that the quark-antiquark potential is double-valued,  however, the upper branch of the potential is unphysical.  It corresponds to an unfavourable string configuration in contrast to the lower branch, which is associated with the lower energy string configuration. Note that at the distance $L_{\phi}=L_{\phi}(r_{m})$ the potential $V_{q\bar{q}}$ tends to be zero and the configuration of two non-interacting straight strings is more preferable energetically. The same holds for $L_{\phi}>L_{\phi}(r_{m})$. Thus, $L_{\phi}(r_{m})$ can be interpreted as the screening length~\cite{Brandhuber:1998bs}. In the work~\cite{Sin:2004yx} it was argued that the time-like screening length, which corresponds to the mean-free path for traveling "light" (gluon) in a medium, has the value $1/\pi T$. As we can see from Fig.~\ref{fig:Vqq-Schwar}{\bf A}, this condition is satisfied for $\theta \gtrsim \pi/9$.

	From Fig.~\ref{fig:Vqq-Schwar}{\bf A} we see that the interquark potential has the Coulomb-like behaviour. If we estimate (\ref{eq:Schwar:Vqq}), we find that $V_{q\bar{q}}$ has the Coulomb-like term. Indeed, the numerical evaluation of the term $I_3$ confirms that at small distances the contribution from $I_{3}$  is inversely proportional to the length $I_3 \sim -1/L_{\phi}$. We show the dependence $I_3$ as a function of $1/L_{\phi}$ in Fig.~\ref{fig:Vqq-Schwar}{\bf B}. Note that $I_{3}$ is double-valued because of the string configuration, see Fig.~\ref{fig:LSchwar-C}{\bf A}. In the deconfined phase the string term vanishes. In our work we are able to approximate eq.\eqref{eq:Schwar:Vqq} as 	
	\begin{equation}\label{eq:Schwar:Vqq:approx}
	V_{q \bar{q}} (L) = - \frac{\kappa}{L_{\phi}} + V_0.
	\end{equation}
	However, in works \cite{Mocsy:2007yj, Dumitru:2007hy} it was suggested that in addition to the Coulomb contribution one has to include the medium-dependent term. 
	In Table~\ref{tab:Vqq-Schwar} we present  values of $\kappa$ and $V_0$, which obtained from fitting of $V_{q\bar{q}}$ in Fig.~\ref{fig:Vqq-Schwar}{\bf A}.
	\begin{table}[!htb]
		\centering
		\begin{tabular}{|c|c|c|c|}
			\hline
			$T_{\rm H}$, GeV & $\theta$ & $\kappa$, GeV$\cdot$fm  & $V_0$, GeV \\
			\hline
			\multirow{3}{*}{0.17}& $\pi/6$ & 0.711459 & 2.67669 \\
			& $\pi/9$& 1.04008 & 2.67669 \\
			& $\pi/12$ & 1.37443 & 2.67669  \\
			\hline
			\multirow{3}{*}{0.20}& $\pi/6$ & 0.707247 & 3.6193  \\
			& $\pi/9$ & 1.03393 & 3.6193 \\
			& $\pi/12$ & 1.3663 & 3.6193 \\
			\hline
			\multirow{3}{*}{0.30}& $\pi/6$ & 0.704129 & 6.08073 \\
			& $\pi/9$ & 1.02937 & 6.08073 \\
			& $\pi/12$ & 1.36027 & 6.08073 \\
			\hline
		\end{tabular}
		\caption{Fitting coefficients of $V_{q\bar{q}}$ ~\eqref{eq:Schwar:Vqq:approx} at temperatures $T_{\rm H}=0.17,0.20,0.30$GeV and angles $\theta=\pi/6,\pi/9,\pi/12$, corresponding to Fig.~\ref{fig:Vqq-Schwar}.}
		\label{tab:Vqq-Schwar}
	\end{table}
	 As one can see, the constant $V_0$ does not depend on the angle $\theta$, but it grows as the temperature $T_{\rm H}$ increases. Surprisingly, at temperature just above the critical one, i.e. $T_{\rm H}\approx0.171\,$GeV, the value of $V_0$  is equal to Euler's number. The Coulomb strength parameter $\kappa$ weakly depends on $T_{\rm H}$ and increases with decreasing angle $\theta$.
	 
\begin{figure}
        \centering
        \includegraphics[width=0.4\linewidth]{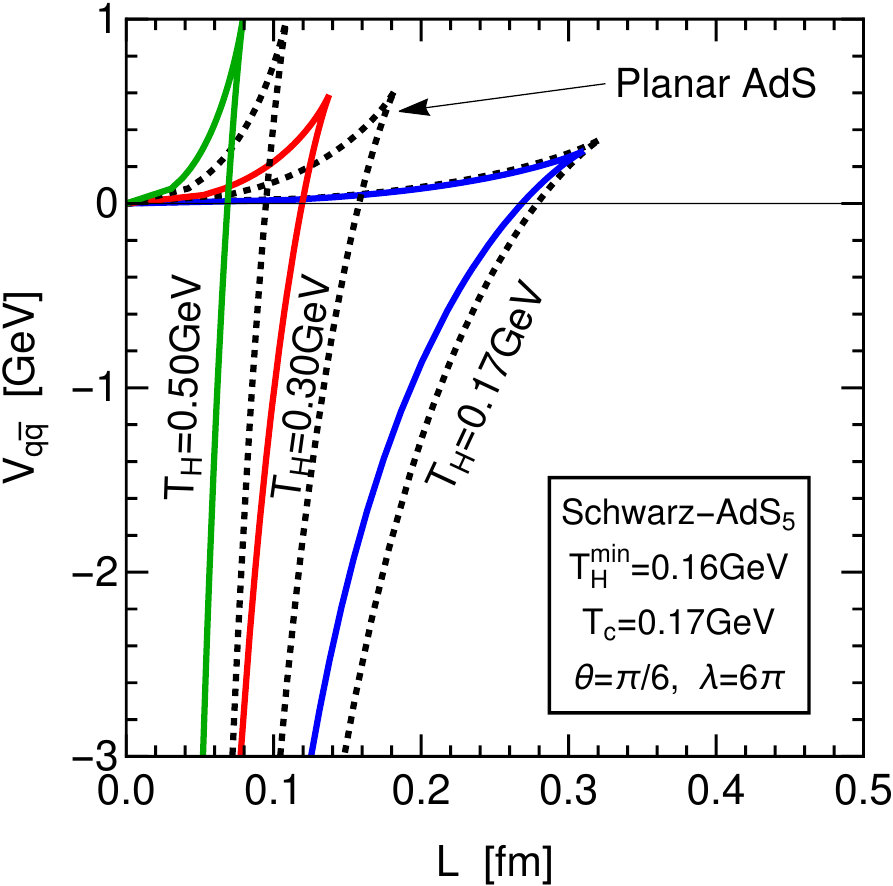}
        \caption{$V_{q\bar{q}}$ in the Schwarzschild-$AdS_5$  (solid curves) and in the planar AdS  (dotted curves)  black holes at fixed $\theta=\pi/6$ and $T_{\rm H}=0.17, 0.30, 0.50\,$GeV.}
        \label{fig:Vqq-Sonnen}
    \end{figure}

	 In Fig.~\ref{fig:Vqq-Sonnen} we compare our results for the potential in the Schwarzschild-$AdS_{5}$ background with $V_{q\bar{q}}$  in the planar AdS black hole~\cite{Brandhuber:1998bs}.
	 For this we set $R=\ell$ in the planar AdS black hole case and fix $\theta= \pi/6$ for the Schwarzschild-$AdS_{5}$ background. We see that for the same quark-antiquark distance $V_{q\bar{q}}$  in Schwarzschild-$AdS_{5}$ (solid curves) takes greater values than the potential in the planar AdS background (dotted curves) ~\cite{Brandhuber:1998bs}. This difference becomes more significant as the temperature increases.
		 
	\subsection{Wilson loop in Kerr-$AdS_5$ black hole}
	Now we turn to the discussion of the holographic Wilson loop in the 5d Kerr-AdS black hole (\ref{eq:metric:asymptAdSKerr-ab})-(\ref{eq:delta:asymptAdSKerr-ab}). 

	For the parametrization of the string  worldsheet  we employ the following gauge condition:
	\begin{equation}\label{eq:asymptAdSKerr-ab:parametrization}
	\tau = T, \qquad \sigma =  \Phi, \qquad y = y(\Phi),\qquad \Phi \in [0, 2\pi L_\Phi].
	\end{equation}
	The components of the induced metric  (\ref{defindm}) are	
	\begin{align}\label{eq:metric:wloop:asymptAdSKerr-ab:induced}
	g_{\tau\tau} &= G_{TT} = -\left(1 + y^2 \ell^{-2} - \frac{2M}{\Delta^3 y^2}\right),\quad g_{\tau\sigma} = G_{T\Phi} = - \frac{2M a \sin^2 \Theta}{ \Delta^3 y^2} \nonumber,\\
	g_{\sigma\sigma} &=G_{\Phi\Phi} + y'^2 G_{yy} =\sin^2 \Theta \left(y^2 + \frac{2M a^2 \sin^2 \Theta}{\Delta^3 y^2} \right) + \frac{y'^2}{1 + y^2 \ell^{-2} - \frac{2M}{\Delta^2 y^2}},
	\end{align}
	where $\Delta$ is given by  (\ref{eq:delta:asymptAdSKerr-ab}) and we denoted $y' \equiv dy/d\Phi$. 	
	We also suppose the following boundary  conditions for the location of the string endpoints
	\be\label{ybndc}
	y\left(-\frac{L_{\Phi}}{2}\right) = y\left(\frac{L_{\Phi}}{2}\right)  =\infty.
	\ee
	Taking into account (\ref{eq:asymptAdSKerr-ab:parametrization})-(\ref{ybndc}) we write down the Nambu-Goto action in the following form:
	\begin{equation}\label{eq:asymptAdSKerr-ab:action}
	S_{\rm NG} =  \frac{T}{2 \pi \alpha'} \int^{\frac{L_{\Phi}}{2}}_{-\frac{L_{\Phi}}{2}} d \Phi\, \sqrt{y'^2 \frac{f_{\Delta^3}(y)}{f_{\Delta^2}(y)} + y^2  F_{\Delta^3}(y) \sin^2 \Theta},
	\end{equation}
	where for clarity we introduced the notation using dimensionless functions
	\begin{equation}
	\begin{aligned}\label{eq:asymptAdSKerr-ab:notations}
	&  f_{\Delta^2}(y) = 1 + y^2 \ell^{-2} - \frac{2M}{\Delta^2 y^2},\qquad  f_{\Delta^3}(y) = 1 + y^2 \ell^{-2} - \frac{2M}{\Delta^3 y^2}\\
	& F_{\Delta^3}(y) = f_{\Delta^3}(y) + \frac{2M a^2 \sin^2 \Theta}{y^4 \Delta^3} \left(1+ y^2\ell^{-2} \right).
	\end{aligned}
	\end{equation}
	
	The system \eqref{eq:asymptAdSKerr-ab:action} has the  integral of motion
	\begin{equation}\label{eq:asymptAdSKerr-ab:hamiltonian}
	\mathcal{H} = - \frac{y^2 F_{\Delta^3}(y) \sin^2 \Theta}{\sqrt{y'^2 \frac{f_{\Delta^3}(y)}{f_{\Delta^2}(y)} + y^2 F_{\Delta^3}(y) \sin^2 \Theta}}.
	\end{equation}
	The turning point is defined by $y' = 0$, so from (\ref{eq:asymptAdSKerr-ab:hamiltonian}) we have
	\be\label{CKerr-def}
	-y \sin\Theta \sqrt{F_{\Delta^3}(y)}\Big\vert_{y=y_m} = - \frac{\ell}{ C},  \quad \textrm{with}\quad y_m = y(\Phi_m).
	\ee
	The equation of motion which follows from \eqref{eq:asymptAdSKerr-ab:hamiltonian} is given by
	\begin{equation}\label{eq:asymptAdSKerr-ab:eom}
	y'^2 = y^2 F_{\Delta^3}(y) \frac{f_{\Delta^2}(y)}{f_{\Delta^3}(y)} \sin^2 \Theta \left[\frac{C^{2}}{\ell^2}\sin^2 \Theta y^2 F_{\Delta^3}(y) - 1 \right] .
	\end{equation}
	Substituting \eqref{eq:asymptAdSKerr-ab:eom} into \eqref{eq:asymptAdSKerr-ab:action} and coming to the integration with respect to $y$ one yields to the  expression:
	\begin{equation}\label{eq:asymptAdSKerr-ab:action-2}
	S_{\rm NG} = \frac{T}{\pi \alpha'} \int^{\infty}_{y_m} d y\,  \frac{C \sin\Theta y \sqrt{F_{\Delta^3}(y)}}{\sqrt{C^2  \sin^2 \Theta y^2 F_{\Delta^3}(y) - \ell^2}} \sqrt{\frac{f_{\Delta^3}(y)}{f_{\Delta^2}(y)}}.
	\end{equation}
	Eq.~\eqref{eq:asymptAdSKerr-ab:action-2} is divergent at the conformal boundary $y \to +\infty$  of the Kerr-$AdS_{5}$ black hole.
	
	Just like in the Schwarzschild-AdS case, the renormalization procedure is a subtraction of the single quarks ``self-energy'', which is represented by the action of a static straight string in Kerr-$AdS_{5}$:
	\begin{equation}\label{contrterm}
	S_0 = \frac{T}{\pi \alpha'} \int_{y_+}^{\infty} dy\, \sqrt{-G_{TT} G_{yy}} = \frac{T}{\pi \alpha'} \left(\int_{y_m}^{\infty} + \int_{y_+}^{y_m} \right) \sqrt{\frac{f_{\Delta^3}(y)}{f_{\Delta^2}(y)}} dy.
	\end{equation}
	Subtracting  \eqref{contrterm} from \eqref{eq:asymptAdSKerr-ab:action-2} we get  
	\begin{equation}\label{eq:asymptAdSKerr-ab:action-RN}
	S_{\rm NG}^{\rm ren} = \frac{T}{\pi \alpha'} \left[ \int^{\infty}_{y_m} d y\, \sqrt{\frac{f_{\Delta^3}(y)}{f_{\Delta^2}(y)}}\left( \frac{C \sin\Theta y \sqrt{F_{\Delta^3}(y)}}{\sqrt{C^2  \sin^2 \Theta y^2 F_{\Delta^3}(y) - \ell^2}} -1\right) - \int^{y_m}_{y_+}dy\sqrt{\frac{f_{\Delta^3}(y)}{f_{\Delta^2}(y)}}\right].
	\end{equation}
	From eq.\eqref{eq:asymptAdSKerr-ab:eom} we find the interquark distance $L_{\Phi}$:
	\begin{equation}\label{eq:asymptAdSKerr-ab:length}
	\frac{L_{\Phi}}{2} =  \int^{\infty}_{y_m} dy\, \frac{\ell}{\sin \Theta y \sqrt{ F_{\Delta^3}(y)} \sqrt{C^{2}\sin^2 \Theta y^2 F_{\Delta^3}(y) - \ell^2}} \sqrt{\frac{f_{\Delta^3}(y)}{f_{\Delta^2}(y)}}.
	\end{equation}
	As in the previous subsection one can find the relation between the string action  \eqref{eq:asymptAdSKerr-ab:action-2} and the quark-antiquark distance  \eqref{eq:asymptAdSKerr-ab:length}. For this reason,  we define \eqref{eq:asymptAdSKerr-ab:action-RN} and \eqref{eq:asymptAdSKerr-ab:length}
	\be\label{eq:asymptAdSKerr-ab:action-chi}
	S_{\rm NG}^{\rm ren} =
	\frac{T}{ \pi \alpha'} I_1(y_m,C),\quad	 \frac{L_{\Phi}}{2} =	I_2(y_m,C).
	\ee
		
	Correspondingly, derivatives of $I_1(y_m,C)$ and  $I_2(y_m,C)$\eqref{eq:asymptAdSKerr-ab:action-chi}  with respect to  $C$ are
	\bea\label{derKerrAdS}
	\frac{\partial I_1(y_m,C)}{\partial C} &= &-  \int^{\infty}_{y_m} d y\, \frac{y\ell^2  \sin\Theta \sqrt{F_{\Delta^3}(y)}}{( \sin^2 \Theta C^2 y^2 F_{\Delta^3}(y) - \ell^{2})^{3/2}} \sqrt{\frac{f_{\Delta^3}(y)}{f_{\Delta^2}(y)}},\\ \label{derKerrAdS2}
	\frac{\partial I_2(y_m,C)}{\partial C} &= &-  \int^{\infty}_{y_m} d y\, \frac{y C\ell \sin\Theta   \sqrt{F_{\Delta^3}(y)}}{( \sin^2 \Theta C^2 y^2 F_{\Delta^3}(y) - \ell^{2})^{3/2}}\sqrt{\frac{f_{\Delta^3}(y)}{f_{\Delta^2}(y)}}.
	\eea
	Comparing (\ref{derKerrAdS}) with (\ref{derKerrAdS2}), we find the following relation 
	\begin{equation}\label{eq:asymptAdSKerr-ab:diffeq}
	\frac{\partial I_2(y_m,C)}{\partial C}=\frac{ C}{\ell} \frac{\partial I_1(y_m,C)}{\partial C} .
	\end{equation}
	Integrating LHS of eq. \eqref{eq:asymptAdSKerr-ab:diffeq}, we obtain
	\be\label{eq:asymptAdSKerr-ab:diffeq-int1}
	\int_0^{C} \frac{\partial I_2(y_m,C)}{\partial C} d C= \frac{L_{\Phi}}{2} + I_{\Delta^3},
	\ee
	where we use the notation
	\be
	I_{\Delta^3}= i\int^{\infty}_{y_m} \frac{dy}{\sin \Theta y \sqrt{ F_{\Delta^3}(y)}} \sqrt{\frac{f_{\Delta^3}(y)}{f_{\Delta^2}(y)}}.
	\ee
	At the same time integrating by parts RHS of eq. \eqref{eq:asymptAdSKerr-ab:diffeq}, we come to
	\be
	\label{eq:asymptAdSKerr-ab:diffeq-int2}
	\int_0^{C} C \frac{\partial I_1(y_m,C)}{\partial C} d C=C I_1(y_m,C) - \int_0^{C} I_1(y_m,C) d C.
	\ee
	The latter integral can be easily  found
	\be\label{eq:asymptAdSKerr-ab:diffeq-int3}
	\begin{aligned}
		&\int_0^{C} I_1(y_m,C) d C\\
		&=\int^{\infty}_{y_m} dy\, \frac{\sqrt{C^2\sin^2 \Theta y^2 F_{\Delta^3}(y) - \ell^2}}{ \sin \Theta y \sqrt{F_{\Delta^3}(y)}} \sqrt{\frac{f_{\Delta^3}(y)}{f_{\Delta^2}(y)}} - C \int^{y_m}_{y_+}dy\sqrt{\frac{f_{\Delta^3}(y)}{f_{\Delta^2}(y)}} - I_{\Delta^3}\ell.
	\end{aligned}
	\ee
	Now collecting  eqs.\eqref{eq:asymptAdSKerr-ab:diffeq}-\eqref{eq:asymptAdSKerr-ab:diffeq-int3}, we get
	\begin{equation}
	S^{\rm ren}_{\rm NG} = \frac{\ell}{C}\frac{T}{\pi \alpha'}\left(\frac{L_{\Phi}}{2} + I_{3}\right),
	\end{equation}
	where
	\begin{equation}\label{eq:Kerr:newint2}
	I_3 = \int^{\infty}_{y_m} dy\, \sqrt{\frac{f_{\Delta^3}(y)}{f_{\Delta^2}(y)}} \left(\frac{\sqrt{ C^2\sin^2 \Theta y^2 F_{\Delta^3}(y) - \ell^2}}{ y \sin \Theta \sqrt{F_{\Delta^3}(y)}} - C\right) - \frac{C}{\ell} \int^{y_m}_{y_+} dy\, \sqrt{\frac{f_{\Delta^3}(y)}{f_{\Delta^2}(y)}}.
	\end{equation}
	So, taking into account \eqref{eq:Kerr:newint2}, we find the same expression as \eqref{eq:Schwar:Vqq} for the quark-antiquark potential
	\be\label{eq:Kerr:Vqq}
	V_{q\bar{q}} = \frac{\sqrt{\lambda}}{\pi \ell^2}y_m \sin \Theta \sqrt{F_{\Delta^3}(y_m)}\left(\frac{L_{\Phi}}{2} + I_3\right).
	\ee
	
\begin{figure}[!htb]
		\centering
		\captionsetup[subfigure]{labelformat=empty}
		\begin{subfigure}{0.485\linewidth}
			\includegraphics[width=1\linewidth]{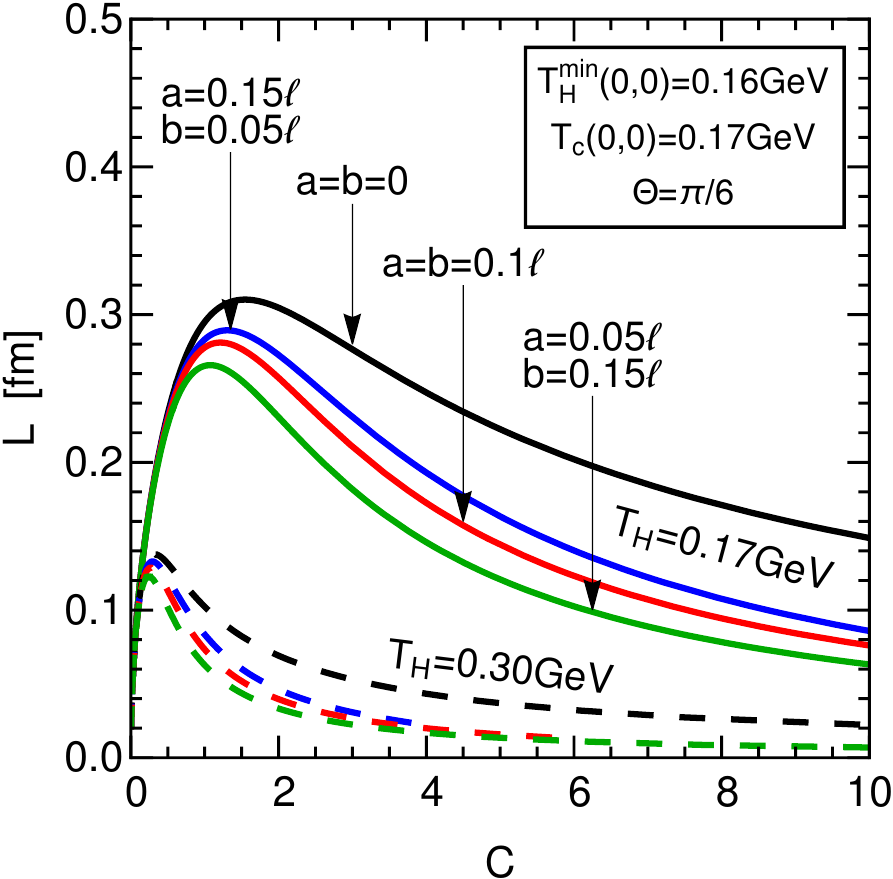}
			\caption{A}
		\end{subfigure}\hfill
		\begin{subfigure}{0.49\linewidth}
			\includegraphics[width=1\linewidth]{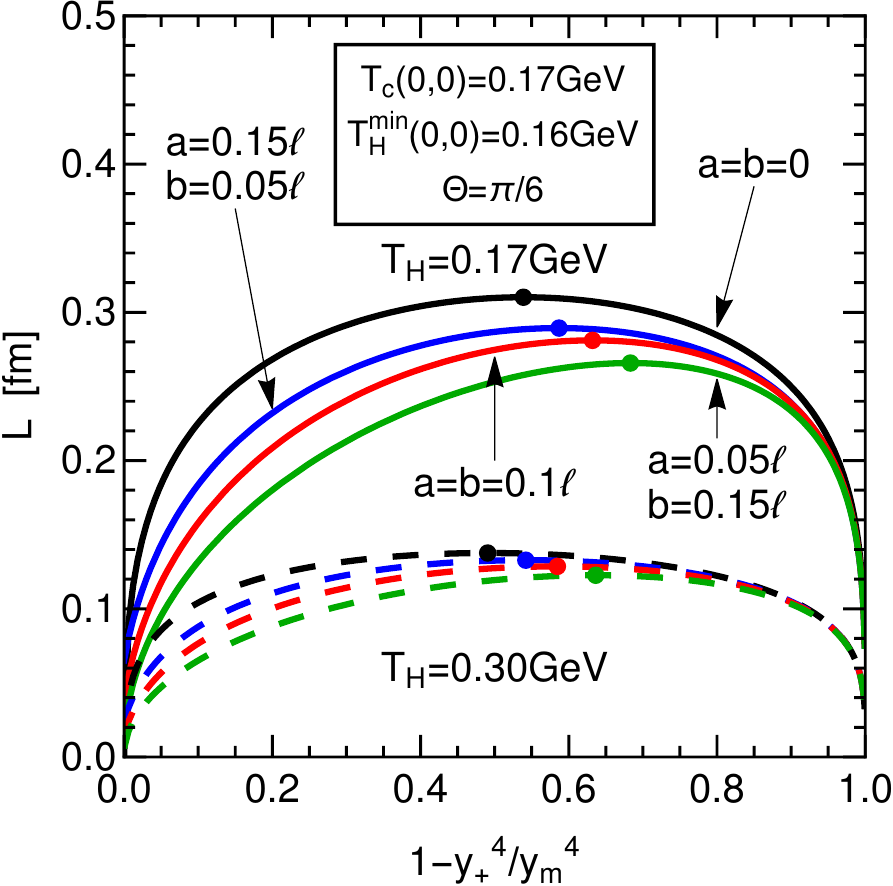}
			\caption{B}
		\end{subfigure}
		\caption{A) The behaviour of the interquark distance  $L_{\Phi}$ on the integration constant $C$~\eqref{CKerr-def}. B)  $L_{\Phi}$ as a function of  $1-\frac{y^4_{+}}{y^{4}_{m}}$.}
		\label{fig:LKerr-C}
	\end{figure}

	Figs.~\ref{fig:LKerr-C}{\bf A}-\ref{fig:LKerr-C}{\bf B} show the dependences of the interquark distance $L_{\Phi}$ (\ref{eq:asymptAdSKerr-ab:length}) on the constant $C$ ~\eqref{CKerr-def}  and the quantity $1-\frac{y_{+}^4}{y_m^4}$, correspondingly. For the plots the temperature $T_{\rm H}$ and the value of the angle $\Theta$ are fixed, while we vary the rotational parameters $a$ and $b$. In both Figs.~\ref{fig:LKerr-C}{\bf A} and {\bf B} we see that $L_{\Phi}$ decreases as the rotational parameters increases, so the interquark distance has the bigger values for zero rotational parameters.
	
\begin{figure}[!htb]
		\centering
		\captionsetup[subfigure]{labelformat=empty}
		\begin{subfigure}{0.49\linewidth}
			\includegraphics[width=0.93\linewidth]{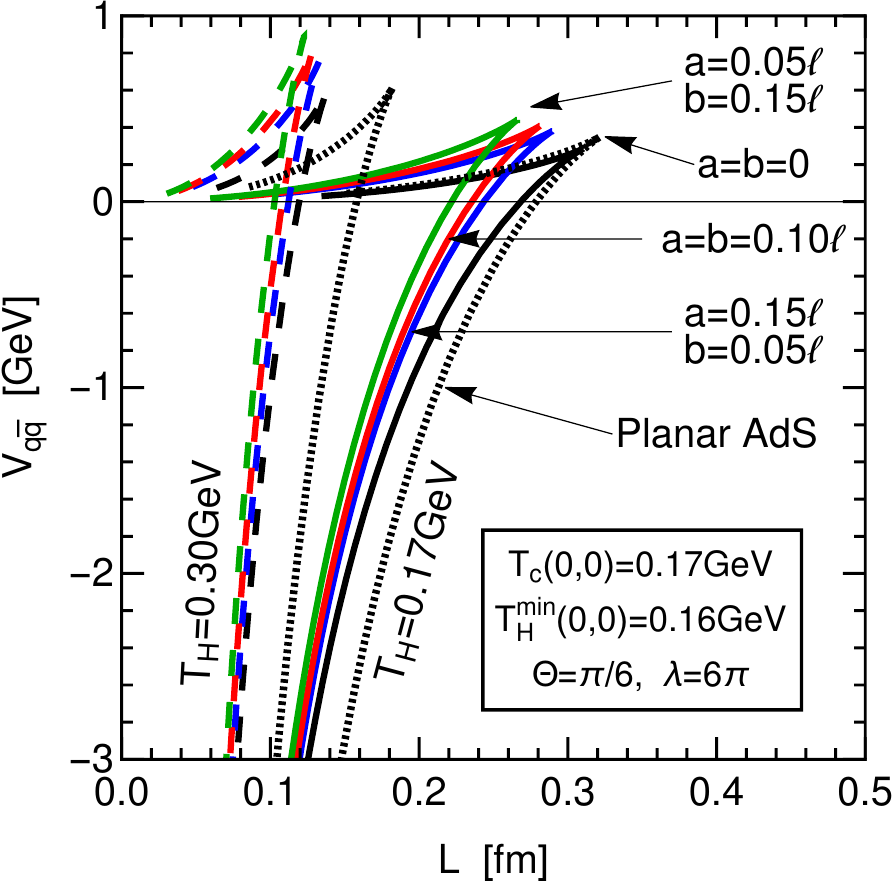}\vspace*{4pt}
			\caption{A}
		\end{subfigure}\hfill
		\begin{subfigure}{0.49\linewidth}
			\includegraphics[width=1\linewidth]{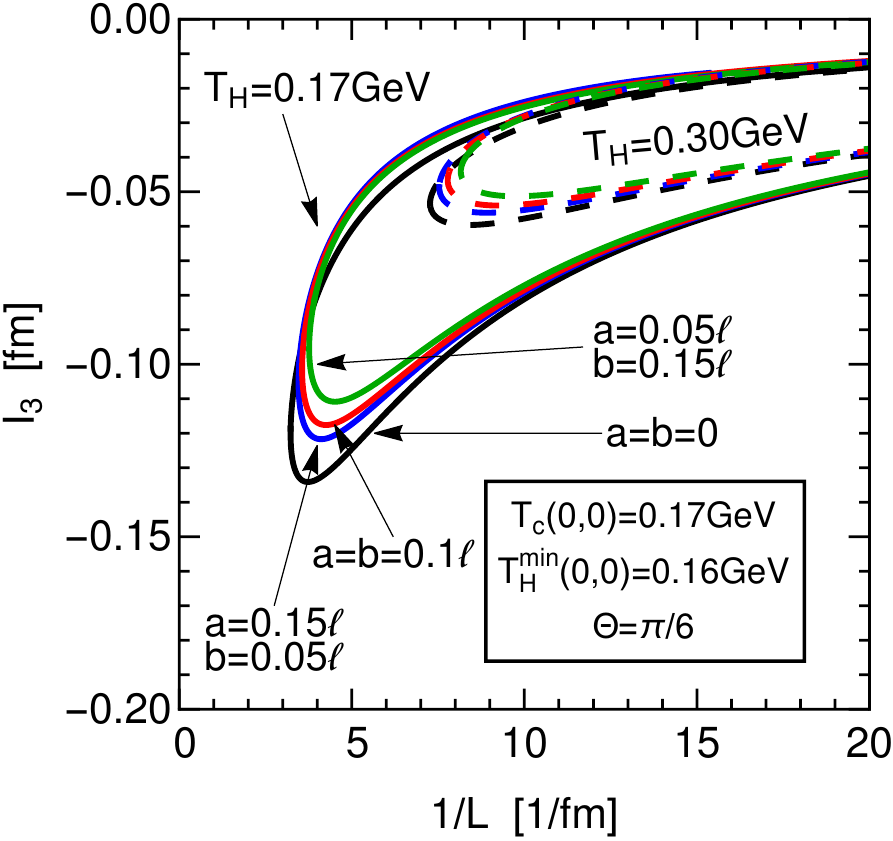}
			\caption{B}
		\end{subfigure}
		\caption{A) The dependence of the quark-antiquark potential $V_{q\bar{q}}$  on the distance $L_{\Phi}$. B) The behaviour of $\mathcal{I}_{3}$ \eqref{eq:Kerr:newint2} on the inverse length $1/L_{\Phi}$.}
		\label{fig:VqqKerr-L}
	\end{figure}

	The dependence of the potential $V_{q\bar{q}}$  on the interquark distance $L_{\Phi}$ is shown in Fig.~\ref{fig:VqqKerr-L}{\bf A}. We set  $T_{\rm H}=0.17$ GeV and $\lambda=6\pi$. From Fig.~\ref{fig:VqqKerr-L}{\bf A} we see that  $V_{q\bar{q}}$ is double-valued, but only the lower curve is significant as for the non-rotating case. This branch corresponds to the string configuration with the lower energy. The potential $V_{q\bar{q}}$ crosses zero at $L_{\Phi,m}=L_{\Phi}(y_{m})$, which can be interpreted as the screening length. The upper branch of the potential, which starts at  $L_{\Phi,m}$, is related to a configuration of two separated straight strings.	
 We can observe in Fig.~\ref{fig:VqqKerr-L}{\bf A} that at $T_{\rm H} =0.17$ GeV the potential $V_{q\bar{q}}$ for non-zero rotational parameters (color solid curves) can have greater values than in  Schwarzschild-$AdS_{5}$  (black solid curve). Increasing the temperature $T_{\rm H} =0.3$ GeV the potential $V_{q\bar{q}}$ in Kerr-$AdS_{5}$ comes closer to $V_{q\bar{q}}$ in the  Schwarzschild-$AdS_{5}$ black hole. It should be noted that  the same values of $V_{q\bar{q}}$ at different temperatures  corresponds to different $L_{\Phi}$, which decreases as the temperature increases.
		
	 From Fig.~\ref{fig:VqqKerr-L}{\bf A} one can see that the potential has also the Coulomb form, which is similar to the Schwarzschild-AdS case. This is also confirmed by the dependence  of the $I_{3}$-term on the inverse interquark distance $1/L$ depicted in Fig.~\ref{fig:VqqKerr-L} {\bf B}. Note that in \cite{Brandhuber:1999jr} it was shown that the quark-antiquark potential in the rotating D3-brane interpolates between the Coulomb and confining parts.
	 
	 We are able to find an approximation of $V_{q\bar{q}}$ (\ref{eq:Kerr:Vqq})  assuming that above the critical temperature $T_{\rm H}=0.17$ GeV it is given  by \eqref{eq:Schwar:Vqq:approx}. We write down $\kappa$ and $V_{0}$ for various values of the rotational parameters $a$ and $b$ in Table~\ref{tab:VqqKerr}.

\begin{table}[!htb]\centering
		\begin{tabular}{|c|c|c|c|c|}
			\hline
			$T_{\rm H}$, GeV & $a/\ell$ & $b/\ell$ & $\kappa$, GeV$\cdot$fm & $V_{0}$, GeV \\
			\hline
			\multirow{4}{*}{0.17} & 0 & 0 & 0.711459 & 2.67669 \\
			& 0.15 & 0.05 & 0.711887 & 2.95381 \\
			& 0.1 & 0.1 & 0.711556 & 3.05421   \\
			& 0.05 & 0.15 & 0.710322 & 3.23494 \\
			\hline
			\multirow{4}{*}{0.30} & 0 & 0 & 0.704129 & 6.08073  \\
			& 0.15 & 0.05 & 0.702162  & 6.39186  \\
			& 0.1 & 0.1 & 0.701980 & 6.62686  \\
			& 0.05 & 0.15 & 0.701468 & 6.96067 \\
			\hline
		\end{tabular}
		\caption{Fitting coefficients in the~\eqref{eq:Schwar:Vqq} at temperatures $T_{\rm  H}=0.17\,$GeV and  $T_{\rm H}=0.30\,$GeV, $\Theta=\pi/6$ and rotational parameters are fixed as in Fig.~\ref{fig:VqqKerr-L}{\bf A}.}
		\label{tab:VqqKerr}
	\end{table}
We see that the Coulomb strength parameter $\kappa$ weakly depends on the rotational parameters (at least for values of $a$ and $b$ under consideration)  and on the temperature. On the contrary, the term $V_0$ strongly depends on the rotation and $T_{\rm H}$.

	\setcounter{equation}{0}
	\section{Jet-quenching parameter} 
	\subsection{Jet quenching parameter in the 5d Schwarzschild-AdS black hole}
	
	In this section, we will discuss the jet-quenching parameter in the 5d Schwarzschild-AdS background  (\ref{Schw.1psi})-(\ref{deltaSchwpsi}) following the holographic prescription.

	To find the jet-quenching parameter $\hat{q}$ we have to come to the  scaled ``light-cone" coordinates
	\be\label{lcSchwpsi}
	\dd x^{+} = \ell^2(\dd t - \ell \dd \phi), \quad \dd x^{-} = \ell^2(\dd t + \ell \dd \phi).
	\ee
	By virtue of the transformations (\ref{lcSchwpsi}) we come to the following form of the Schwarzschild-$AdS_{5}$ metric (\ref{Schw.1psi})
	\begin{align}\label{Schwarzlcpsi}
	ds^2 &= \frac{1}{4 \ell^4} \left( \frac{r^2}{\ell^2} \sin^2 \theta - \frac{f(r)}{r^2} \right) \left[ (dx^-)^2 + (dx^+)^2 \right] - \frac{1}{2 \ell^4} \left( \frac{r^2}{\ell^2} \sin^2 \theta + \frac{f(r)}{r^2} \right) dx^- dx^+ \nonumber\\
	&+ \frac{r^2}{f(r)} dr^2 + r^2 d\theta^2 + r^2 \cos^2 \theta d\psi^2,
	\end{align}
	where $f(r)$ is given by (\ref{deltaSchwpsi}).
	We have to study  a holographic  light-like  Wilson loop (\ref{eq:JQmain}) in the background (\ref{Schwarzlcpsi}). We choose
	the coordinates on the string worldsheet as follows
	\be\label{parampsi}
	\tau = x^{-}, \quad \sigma = \psi.
	\ee
	Moreover, for the string configuration we also suppose
	\be\label{param2psi}
	x^{\mu} = x^{\mu}(\sigma), \quad \theta(\sigma) = \const, \quad x^{+}(\sigma) = \const.
	\ee
	Taking into account (\ref{parampsi})-(\ref{param2psi}) we find 
	the corresponding Nambu-Goto action (\ref{defNGact})
	\be\label{eq:schwarz-actionpsi}
	S = \frac{L^-}{2 \pi \alpha'} \int^{L/2}_{-L/2} d \psi\, \frac{r}{2\ell^2} \sqrt{\left( \frac{f(r)}{r^2}  -\ell^{-2}r^2 \sin^2 \theta \right) \left( \cos^2 \theta + \frac{r'^2}{f(r)}\right)},
	\ee
	where we define $r'\equiv\partial r/ \partial \psi$.
	The corresponding first integral  is given by
	\be\label{FISchwarzpsi}
	\mathcal{H} = - \frac{\cos^2 \theta \sqrt{f(r) -r^4  \ell^{-2}\sin^2 \theta }}{2 \ell^2 \sqrt{\cos^2 \theta + \frac{r'^2}{f(r)}}}.
	\ee
	From (\ref{FISchwarzpsi}) we find the equation of motion 
	\be\label{eq:schwarz-eompsi}
	r'^2 = \frac{f(r) \cos^2 \theta}{4 C^2 \ell^6} [\cos^2 \theta ( f(r) \ell^2- r^4 \sin^2 \theta ) - 4C^2 \ell^6],
	\ee
	where $C$ is some constant.
	Plugging \eqref{eq:schwarz-eompsi} into \eqref{eq:schwarz-actionpsi} and coming to the integration with respect to $r$ the Nambu-Goto action can be represented as
	\be\label{eq:schwarz-action2psi}
	S = \frac{L^-}{ \pi \alpha'} \int^{\infty}_{r_{h} + \epsilon} dr\, \frac{\cos \theta (f(r) \ell^2- r^4 \sin^2 \theta)}{2 \ell^3 \sqrt{f(r)} \sqrt{ \cos^2 \theta ( f(r) \ell^2 - r^4 \sin^2 \theta) - 4 C^2 \ell^6}}.
	\ee
	The relation (\ref{eq:schwarz-action2psi}) is divergent on the background boundary $r\to + \infty$  and has to be renormalized. Moreover, since eq.(\ref{eq:schwarz-action2psi}) contains a multiplier $f(r)^{-1/2}$
	we regularize the action on the lower bound as $r_h + \epsilon$.
	The normalization of eq.(\ref{eq:schwarz-action2psi}) can be performed through the subtraction of the static mass of the quark and antiquark, which is given by
	\be\label{eq:schwarz-action3psi}
	S_0 = \frac{ L^-}{ \pi \alpha'} \int^{\infty}_{r_{h} + \epsilon} d r\, \frac{\sqrt{f(r) \ell^2 - r^4 \sin^2 \theta }}{2 \ell^3 \sqrt{f(r)}}.
	\ee
	With (\ref{eq:schwarz-action3psi}) the regularized string action is given by
	\bea\label{SregSchwpsi}
	S^{\rm reg} & =& S - S_0 \\
	&=& \frac{ L^-}{ \pi \alpha'} \int^{\infty}_{r_{h} +\epsilon} d r\, \frac{\sqrt{f(r) \ell^2 - r^4 \sin^2 \theta }}{2 \ell^3 \sqrt{f(r)}} \left( \frac{\cos \theta \sqrt{ f(r) \ell^2 - r^4 \sin^2 \theta}}{\sqrt{ \cos^2 \theta ( f(r) \ell^2- r^4 \sin^2 \theta ) - 4 C^2 \ell^6}} - 1 \right).\nonumber
	\eea
	Expanding (\ref{SregSchwpsi}) for small $C$ (in the low energy limit) we  find
	\be\label{Schw-actCpsi}
	S^{\rm reg} = \frac{L^{-}}{\pi \alpha'}\frac{\ell^2 C^2}{\cos^2\theta}\mathcal{I},
	\ee
	where we denote by $\mathcal{I}$ the following integral
	\be
	\mathcal{I} =\int^{\infty}_{r_m}\frac{dr}{\sqrt{f(r)}\sqrt{f(r)-r^{4}\ell^{-2}\sin^{2}\theta}}
	\ee
	and $r_{m}$ is defined as a positive real solution to the equation
	\be
	r^2 + r^{4}\ell^{-2}\cos^2\theta -2M =0.
	\ee
	
	To find the relation between $L$ and $C$ we remember that $r(\pm L/2)=\infty$ and we have:
	\be\label{LC-schwarzpsi}
	\frac{L}{2} =  \frac{2 C \ell^3}{\cos \theta} \int^{\infty}_{r_h} \frac{dr}{\sqrt{f(r)} \sqrt{\cos^2 \theta(f(r) \ell^2 - r^4 \sin^2 \theta) - 4 C^2 \ell^6}},
	\ee
	or for small  $C$ we get
	\be\label{LC-schwarz2psi}
	\frac{L}{2} = \frac{2\ell^2 C}{\cos^2\theta}\mathcal{I}.
	\ee
	
	Deriving $C$ from (\ref{LC-schwarz2psi}) and substituting into the action (\ref{Schw-actCpsi}) we come to
	\be\label{Sreg-schwarz}
	S^{\rm reg}= \frac{L^{-}}{\pi \alpha'}\frac{L^2\cos^2\theta}{16\ell^2 \int^{\infty}_{r_m}\frac{dr}{\sqrt{f(r)}\sqrt{f(r)- r^{4}\ell^{-2}\sin^{2}\theta}}}.
	\ee
	We note that $r_{m}\geq r_h$ and $r_m$ coincides  with $r_h$ only for $\theta=0$. In this case we need to shift the turning point  to regularize the divergence near $r_h$, i.e. $r_{m}|_{\theta=0} = r_{h}+\epsilon$.
	
	Taking into account \eqref{eq:JQmain3} and \eqref{Sreg-schwarz}, we find the jet-quenching parameter
	\be\label{jetqSchw}
	\hat{q} = \frac{\sqrt{\lambda}}{\sqrt{2}\pi}  \frac{\cos^{2}\theta}{\ell^{4}\int^{\infty}_{r_m}\frac{dr}{\sqrt{f(r)}\sqrt{f(r)- r^{4}\ell^{-2}\sin^{2}\theta}}},
	\ee
	with $\lambda = \ell^4/\alpha'^2$ and  $f(r)$ given by (\ref{deltaSchwpsi}). 
	
	We are not able to calculate (\ref{jetqSchw}) analytically.
	However, for small $\theta$ and $\ell=1$ we can estimate the  expression for $\hat{q}$. First, we find the integral in the denominator of (\ref{jetqSchw}) for  $\ell=1$ and small $\theta$
	{\footnotesize{
			\bea
			&&\int^{\infty}_{r_{h}+\epsilon}\frac{dr}{r^{4}+r^{2}-2M} = \left(\frac{\ln\left(\frac{2r-\sqrt{2}\sqrt{\sqrt{8M+ 1}- 1}}{2r+\sqrt{2}\sqrt{\sqrt{8M+ 1}- 1}}\right)}{\sqrt{2}\sqrt{8M+ 1}\sqrt{\sqrt{8M+ 1}- 1}}-\frac{\sqrt{2}\arctan\left(\frac{\sqrt{2}r}{\sqrt{\sqrt{8M+ 1}+ 1}}\right)}{\sqrt{8M+ 1}\sqrt{\sqrt{8M+ 1}+ 1}}\right)\Big|^{\infty}_{r_{h}+\epsilon}\\
			&&= \frac{1}{(1+2r^2_{h})\sqrt{1+r^2_{h}}}\left(\arctan(\frac{r_{h}}{\sqrt{r^2_{h}+1}}) -\frac{\pi}{2} - \frac{\ln\left(\frac{\epsilon}{\epsilon+2r_{h}}\right)\sqrt{r^2_{h}+1}}{2r_{h}} + \frac{\ln(\infty)\sqrt{r^2_{h}+1}}{2r_{h}}\right)\nonumber\\
			&&=\frac{\sqrt{2}}{(\pi^2 T^2_{\rm H}+\pi T_{\rm H}\sqrt{\pi^2 T^2_{\rm H}-2})\sqrt{\pi T_{\rm H}(\sqrt{\pi^2 T^2_{\rm H}-2}+\pi T_{\rm H})+1}}\times\nonumber\\
			&&\times\Bigl(-\textrm{arccot}\Bigl(\frac{\pi T_{\rm H}+\sqrt{\pi^2 T^2_{\rm H} -2}}{\sqrt{2}\sqrt{\pi T_{\rm H}\left(\sqrt{\pi^2 T^2_{\rm H}-2} + \pi T_{\rm H}\right) +1}}\Bigr)  +\frac{\sqrt{\pi  T_{\rm H} (\sqrt{\pi ^2 T_{\rm H}^2-2}+\pi T_{\rm H})+1}(\ln(\infty)-\ln(\frac{\epsilon}{\epsilon +2r_{h}}))}{\sqrt{2}(\pi T_{\rm H} + \sqrt{\pi^2 T^2_{\rm H} -2})}\Bigr).\nonumber
			\eea}}
	
	Thus, for  high temperatures, $\ell=1$ and small $\theta$ we get the dependence $\hat{q}$ on the temperature as  for the planar AdS black brane \cite{LRW}
	\be\label{ApproxJQSchwA}
	\hat{q} \sim \frac{\sqrt{\lambda}}{2\pi} \kappa T^{3}_{\rm H},
	\ee
	where $\kappa$ is some constant.
	
\begin{figure}
		\centering
		\captionsetup[subfigure]{labelformat=empty}
		\begin{subfigure}{0.49\linewidth}
			\includegraphics[width=1\linewidth]{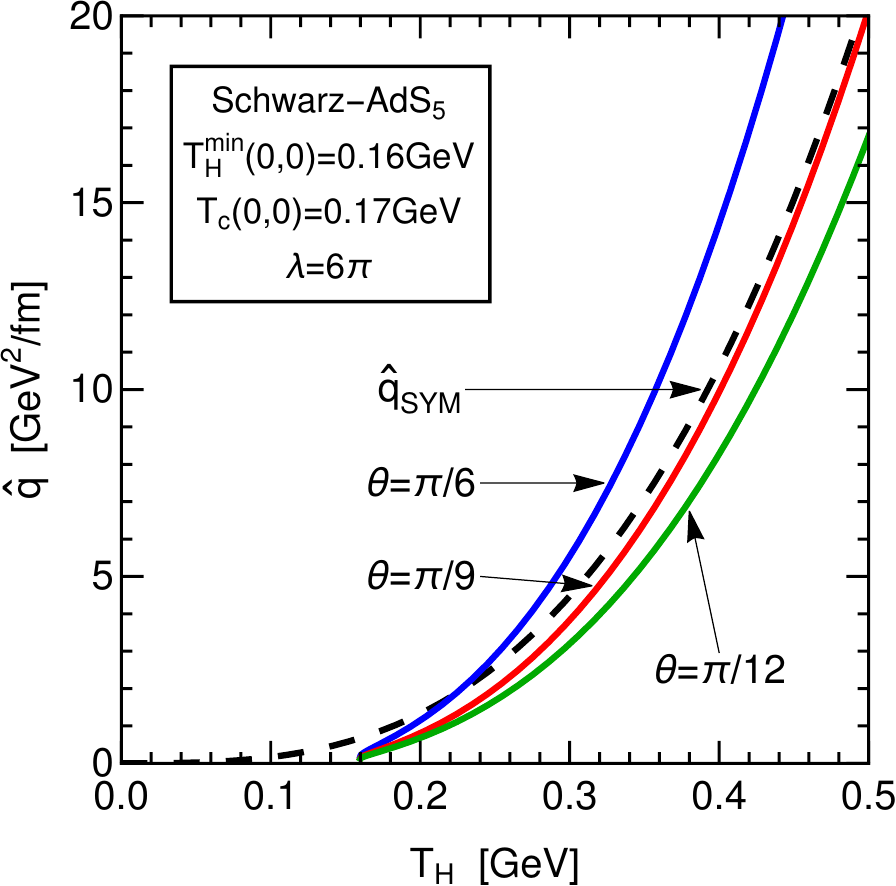}
			\caption{A}
		\end{subfigure}\hfill
		\begin{subfigure}{0.49\linewidth}
			\includegraphics[width=0.97\linewidth]{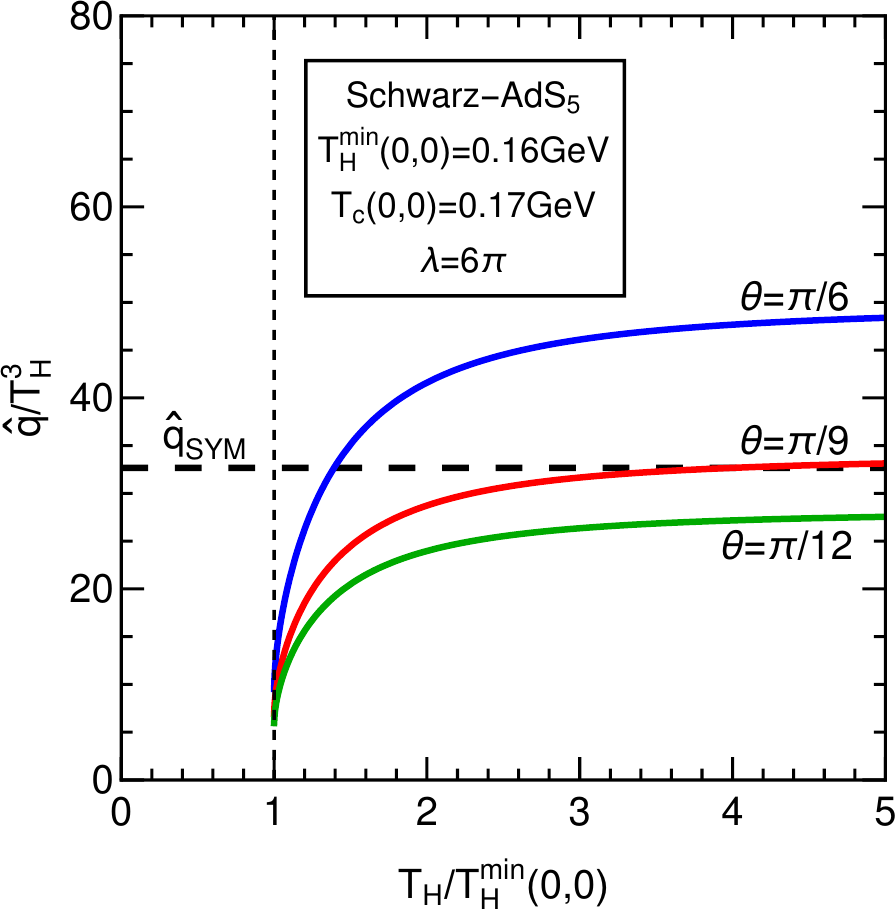}
			\caption{B}
		\end{subfigure}
		\caption{A) The dependence of $\hat{q}$ on the temperature $T_{\rm H}$ for different values of  $\theta$. The case  $\hat{q}_{\rm SYM}$ for the planar $AdS_{5}$ black brane (see eq.~\eqref{JQintro}) is shown by the black dashed curve. B) $\hat{q}/T^3_{\rm H}$ as a function of $T_{\rm H}/T^{\rm min}_{\rm H}$ for different values of $\theta$. The case  $\hat{q}_{\rm SYM}$ for the planar $AdS_{5}$ black brane is shown by the black dashed line.}
		\label{fig:qhat-T-Schwarz}
	\end{figure}

	In Fig.~\ref{fig:qhat-T-Schwarz}{\bf A} we show the jet-quenching parameter $\hat{q}$ \eqref{jetqSchw} for different $\theta$ (solid curves) as a function of the temperature $T_{\rm H}$.  
	
	Note that we are able to trace the behaviour of $\hat{q}$ starting from the $T^{\rm min}_{\rm H}$, since below this temperature the black hole doesn't exist. 
	From  Fig.~\ref{fig:qhat-T-Schwarz}{\bf A} we see that for small values of the angle $\theta$, the parameter $\hat{q}$ is quite close to the curve $\hat{q}_{\rm SYM}$ (see eq.~\eqref{JQintro}) corresponding to the planar $AdS_{5}$ black brane from \cite{LRW}. It is interesting that we can find a small value of $\theta$, for instance $\theta < \pi/9$, such that the jet-quenching parameter $\hat{q}$ is even smaller than $\hat{q}_{\rm SYM}$ for all values of $T_{\rm H}$. Thus, we see that the value of the jet-quenching parameter for the Schwarzschild-AdS black hole depends on the location of the quark.
	
	The fact that $\hat{q}$ in the Schwarzschild-AdS background for very small angles $\theta$ can lie below the curve  $\hat{q}_{\rm SYM}$ in the AdS black brane background is explained by that we have the different contributions of the metric coefficients for the $AdS_5$ black holes with planar  and spherical horizons. Taking $\theta \sim 0$ we change the contribution from the $G_{x^{-}x^{-}}$ term (\ref{Schwarzlcpsi}) in the Nambu-Goto action (\ref{eq:schwarz-actionpsi}).

	It is  instructive to look on the dependence of  $\hat{q}/T^3_{\rm H}$ on the ratio $T_{\rm H}/T^{\rm min}_{\rm H}$. We depict this for the Schwarzschild-AdS black hole (solid curves) and AdS black brane (dashed curve) in Fig.~\ref{fig:qhat-T-Schwarz}{\bf B}.
 Comparing to $\hat{q}_{\rm SYM}/T^3_{\rm H}$, which is constant for all range of $T_{\rm H}/T^{\rm min}_{\rm H}$, the quantity  $\hat{q}/T^3_{\rm H}$ for the Schwarzschild-$AdS_5$ background has a nonlinear behaviour on  $T_{\rm H}$  up to some value of $T_{\rm H}$,  above which it also takes a constant value. From this figure, one can conclude that at high temperature the jet-quenching parameter $\hat{q}$ in the AdS black hole has a generic dependence on $T_{\rm H}$ as $T^3_{\rm H}$.
	
	\subsection{Jet quenching parameter in the Kerr-$AdS_{5}$ background}
	Now we turn to the calculation of the jet-quenching parameter $\hat{q}$ in the Kerr-$AdS_{5}$ background  (\ref{eq:metric:asymptAdSKerr-ab}) with two arbitrary rotational parameters.
	Here we use the following ``light-cone'' coordinates suggested in  \cite{Cvetic:2005nc}
	\begin{align}\label{lightcone}
	\dd x^+ = \dd T - a \dd \Phi, \qquad \dd x^- = \dd T + a \dd \Phi.
	\end{align}
	Taking into account (\ref{lightcone}) and putting for simplicity $\ell =1$  the Kerr-$AdS_{5}$ metric  \eqref{eq:metric:asymptAdSKerr-ab}   takes the  form:
	\begin{equation}
	\begin{aligned}\label{asymptAdS-metric-lightcone:2}
	ds^{2} \simeq &\frac{1}{4} \zeta(y) (\dd x^-)^2 + \frac{1}{4} \left(\eta(y) + \frac{2M}{\Delta^3 y^2} (1 + \sin^2 \Theta)^2 \right) (\dd x^+)^2\\
	&+ \frac{1}{2} \left(\xi(y) + \frac{2M}{\Delta^3 y^2} (1 - \sin^4 \Theta) \right) \dd x^- \dd x^+ - \frac{2M}{\Delta^3 y^2} b (1 - \sin^4 \Theta)\, \dd \Psi \dd x^+\\
	&- \frac{2M}{\Delta^3 y^2} b \cos^4 \Theta\, \dd \Psi \dd x^- + \cos^2 \Theta \left( y^2 + \frac{2M}{\Delta^3 y^2} b^2 \cos^2 \Theta \right) \dd \Psi^2\\
	&+ \frac{\dd y^2}{f_{\Delta^2}(y)} + y^2 \dd \Theta^2,
	\end{aligned}
	\end{equation}
	where $f_{\Delta^2}(y)$ is given by \eqref{eq:asymptAdSKerr-ab:notations}, and we  introduced the following notation:
	\begin{equation}\label{etafdef}
	\begin{aligned}
	\eta(y) &= 1 + y^2 - \frac{y^2}{a^2}\sin^2 \Theta,\quad \xi(y) = - (1 + y^2) - \frac{y^2}{a^2 }\sin^2 \Theta,\\
	\zeta(y) &= \eta(y) - \frac{2M}{\Delta^3 y^2}\cos^4 \Theta.
	\end{aligned}
	\end{equation}

	We parametrize the string worldsheet as follows
	\be\label{eq::jq:kerr:ads:parampsi:2}
	\tau = x^{-}, \qquad \sigma = \Psi,
	\ee
	so $L$ is a length along $\Psi$ and we have $L^{-}$ along the light-cone direction. 
	We also suppose that 
	\be\label{eq:jq:kerr:ads:param2psi:2}
	x^{\mu} =x^{\mu}(\sigma),
	\ee
	thus the Wilson loop lies at constant $x^{+}$ and $\Theta$ 
	\be
	\Theta(\sigma) = \const, \qquad x^{+}(\sigma) = \const.
	\ee
	We also impose the following constraint  for the string endpoints
	\be
	y\Bigl(\pm  L/2\Bigr) = \infty,
	\ee
	and 
	\be
	y(\sigma) = y(-\sigma).
	\ee
	

	The string dynamics in the background (\ref{asymptAdS-metric-lightcone:2}) is governed by the Nambu-Goto action (\ref{defNGact}) defined as
	\begin{equation}\label{eq:jq:kerr:ads:action:2}
	S = \frac{L^-}{2\pi \alpha'} \int^{\Psi'}_0 d \Psi\, \frac{1}{2\ell^2} \sqrt{ \frac{y'^2 \zeta(y)}{f_{\Delta^2}(y)} + \beta(y)},
	\end{equation}
	where for simplicity we introduce
	\begin{equation}\label{betafdef}
	\beta(y) = \cos^2 \Theta (\eta(y) \frac{2M}{\Delta^3 y^2} b^2 \cos^2 \Theta + \zeta(y) y^2)
	\end{equation}
	and $y' \equiv \frac{d y}{d \Psi}$.
	The integral of motion which follows from (\ref{eq:jq:kerr:ads:action:2}) is given by
	\begin{equation}\label{eq:jq:kerr:ads:intm}
	\mathcal{H} =\frac{\beta(y)}{2 \sqrt{\beta(y) + \frac{y'^2 \zeta(y)}{f_{\Delta^2}(y)}}}.
	\end{equation}
	From (\ref{eq:jq:kerr:ads:intm}) one obtains  the equation of motion
	\begin{equation}\label{eq:jq:kerr:ads:eom:2}
	y'^2 =\frac{f_{\Delta^2}(y) \beta(y)}{\zeta(y)} \left(\frac{\beta(y)}{4 C^2} - 1\right).
	\end{equation}
	By owning \eqref{eq:jq:kerr:ads:eom:2}, we find the Nambu-Goto action \eqref{eq:jq:kerr:ads:action:2}  in terms of  the holographic coordinate $y$ 
	\begin{equation}\label{eq:jq:kerr:action:nreg}
	S = \frac{L^-}{\pi \alpha'} \int^{\infty}_{y_+} dy\, \frac{\sqrt{\zeta(y) \beta(y)}}{2 \sqrt{f_{\Delta^2}(y)(\beta(y) -  4 C^2)}}.
	\end{equation}
	As in the non-rotating case considered in the previous section  the string action eq.~(\ref{eq:jq:kerr:action:nreg}) has a divergence near the boundary $y\to + \infty$. To renormalize it one has to  subtract the ``self-energy''  of two quarks, i.e. the action of two straight strings  in the background  (\ref{asymptAdS-metric-lightcone:2}):
	\be
	S_0 = \frac{L^-}{\pi \alpha'} \int^{\infty}_{y_+} dy\, \sqrt{|G_{x^-x^-}G_{yy}|} =  \frac{L^-}{\pi \alpha'} \int^{\infty}_{y_+} \frac{\sqrt{\zeta(y)}}{2 \sqrt{f_{\Delta^2}(y)}}.
	\ee
	The regularized Nambu-Goto action ~(\ref{eq:jq:kerr:action:nreg})  is given by
	\begin{equation}\label{eq:jq:kerr:ads:action:reg:2}
	S^{\rm reg} = S - S_0 = \frac{L^-}{\pi \alpha'} \int^{\infty}_{y_+} dy\, \frac{\sqrt{\zeta(y)}}{2 \sqrt{f_{\Delta^2}(y)}} \left(\frac{\sqrt{\beta(y)}}{ \sqrt{\beta(y) - 4 C^2}} - 1\right).
	\end{equation}
	
	To find a relation between the constant $C$ and the interquark distance $L$, we can use the relation \eqref{eq:jq:kerr:ads:eom:2}:
	\begin{equation}\label{eq:jq:kerr:ads:lenght:2}
	\frac{L}{2} =  \int^{\infty}_{y_+} dy\, \frac{2 C \sqrt{\zeta(y)}}{\sqrt{f_{\Delta^2}(y) \beta(y)} \sqrt{\beta(y) - 4 C^2}}.
	\end{equation}
	In the limit for small $C$  the Nambu-Goto action \eqref{eq:jq:kerr:ads:action:reg:2} and the distance $L$ between string endpoints \eqref{eq:jq:kerr:ads:lenght:2} take the following form, correspondingly
	\bea\label{eq:jq:kerr:ads:series-ab:2}
	&& S^{\rm reg} =  \frac{L^-}{\pi \alpha'} C^2 \mathcal{I} + \mathcal{O}(C^4), \,\,\,\,\,\,\\ \label{eq:jq:kerr:ads:series-ab:3}
	&& \frac{L}{2} = 2C \mathcal{I} + \mathcal{O}(C^3),
	\eea
	where for convenience we introduce 
	\be\label{eq:subint:CLim:a}
	\mathcal{I} = \int^{\infty}_{y_+} dy\, \frac{ \sqrt{\zeta(y)}}{\beta(y) \sqrt{f_{\Delta^2}(y)}},
		\ee
		with $f_{\Delta^2}(y)$, $\zeta(y)$ and $\beta(y)$ are given by \eqref{eq:asymptAdSKerr-ab:notations}, \eqref{etafdef} and \eqref{betafdef}, correspondingly.
	Finally, plugging (\ref{eq:jq:kerr:ads:series-ab:3}) into (\ref{eq:jq:kerr:ads:series-ab:2}) we find that the Nambu-Goto action is
	\be\label{Sreg:Kmath2}
	S^{\rm reg}= \frac{L^{-}}{\pi \alpha'}\frac{L^2}{16 \mathcal{I}}.
	\ee
	Correspondingly,  by owning (\ref{eq:JQmain3}) the jet-quenching parameter  $\hat{q}$ in the  Kerr-$AdS_{5}$ background can be read off as follows
	\be\label{hatqKerr}
	\hat{q} = \frac{\sqrt{\lambda}}{\sqrt{2} \pi \mathcal{I}},
	\ee
	or restoring the dimension with  $\ell$ one can write $\hat{q} = \frac{\sqrt{\lambda}}{\sqrt{2} \pi \ell^4 \mathcal{I}}$.
	
\begin{figure}
		\centering
		\captionsetup[subfigure]{labelformat=empty}
		\begin{subfigure}{0.48\linewidth}
			\includegraphics[width=1\linewidth]{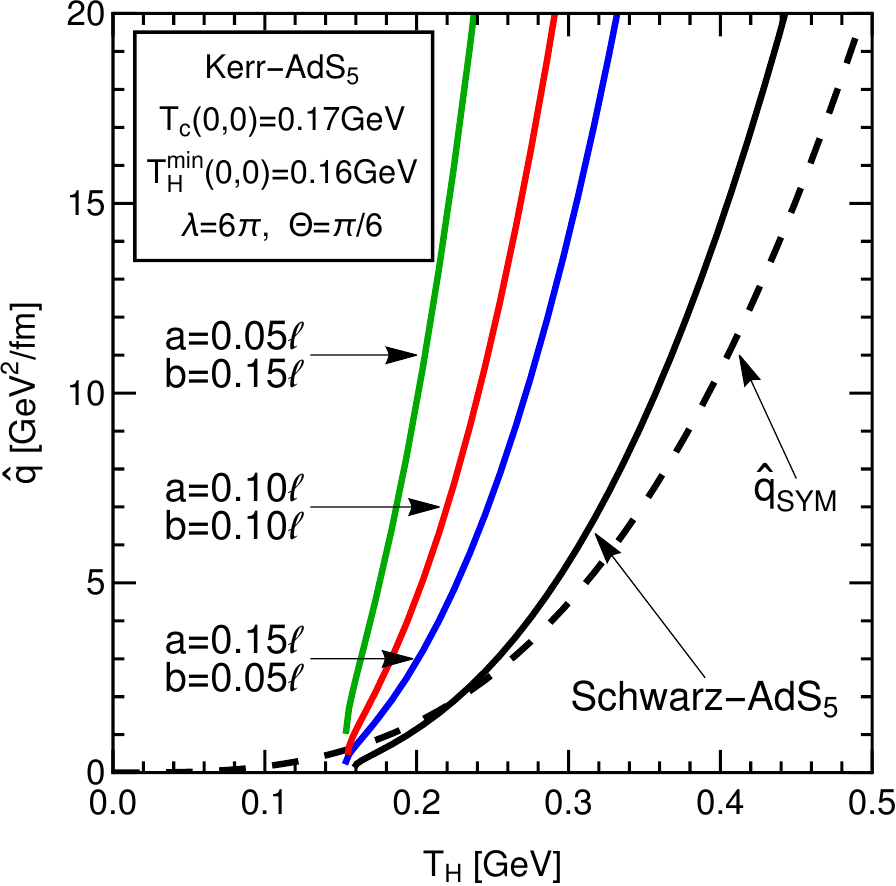}
			\caption{A}
		\end{subfigure}\hfill
		\begin{subfigure}{0.48\linewidth}
			\includegraphics[width=1\linewidth]{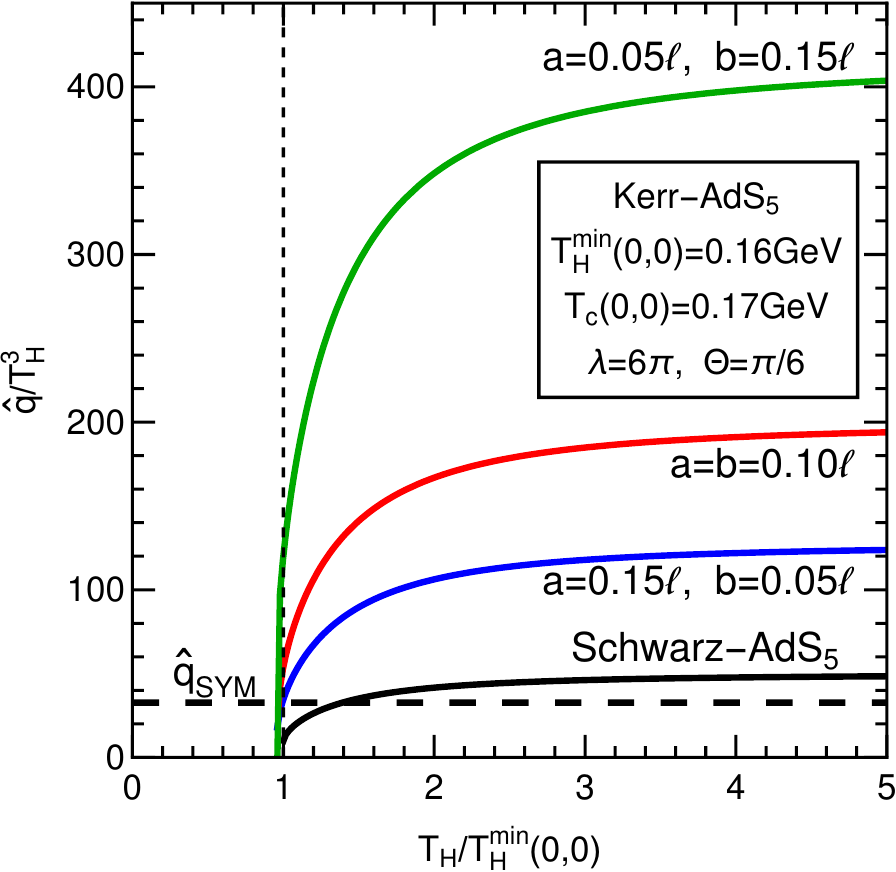}
			\caption{B}
		\end{subfigure}
		\caption{A) The jet-quenching parameter $\hat{q}$ as a function of $T_{\rm H}$ in the Kerr-AdS$_5$ geometry at different rotational parameters (color solid curves), in  the AdS black hole with a planar horizon (the dashed curve) and in the Schwarzschild-$AdS_5$ background (the solid black curve). B) $\hat{q}/T^3_{\rm H}$ as a function of $T_{\rm H}/T^{\rm min}_{\rm H}$ for different values of rotational parameters.}
		\label{fig:qhat-T-Kerr}
	\end{figure}

	In  Fig.~\ref{fig:qhat-T-Kerr}{\bf A} we show the behaviour of the jet-quenching parameter $\hat{q}$ in the Kerr-$AdS_{5}$ background (\ref{hatqKerr}) as a function of $T_{\rm H}$ for different rotational parameters $a$ and $b$ (color solid curves). From this figure we find that  non-zero rotational parameters increase  $\hat{q}$ comparing to $\hat{q}_{\rm SYM}$, which is calculated in the planar AdS black brane background. However, it worth to be noted that for  $a<b$ the value of the jet-quenching parameter $\hat{q}$ is greater than for  $a\geq b$.
	We see that  $\hat{q}$ increases with the temperature faster in the Kerr-AdS background then in the non-rotating AdS black hole, so one can say that the rotation promotes to the energy loss.

	The dependence  $\hat{q}/T^{3}_{\rm H}$  on $T_{\rm H}/T_{\rm H}^{\rm min}(0,0)$ for the Kerr-$AdS_{5}$ background is depicted in Fig.~\ref{fig:qhat-T-Kerr}{\bf B}. 
	We observe that $\hat{q}/T^{3}_{\rm H}$ increases  up to some value of $T_{\rm H}$,
	above which it becomes to be constant similar for the case of the Schwarzschild-AdS black hole. Therefore, at high temperatures the jet-quenching parameter $\hat{q}$ is proportional to $T^{3}_{\rm H}$ as in the Schwarzschild-$AdS_{5}$ (\ref{ApproxJQSchwA}) and the planar cases (\ref{JQintro}) \cite{LRW}  
	\be
	\hat{q} \sim  \kappa_{\rm rot} T^{3}_{\rm H},
	\ee
	where the coefficient $\kappa_{\rm rot}$ depends on values of $a$ and $b$.

	\begin{figure}[h!]
		\centering
		\captionsetup[subfigure]{labelformat=empty}
		\begin{subfigure}{0.49\linewidth}
			\includegraphics[width=0.99\linewidth]{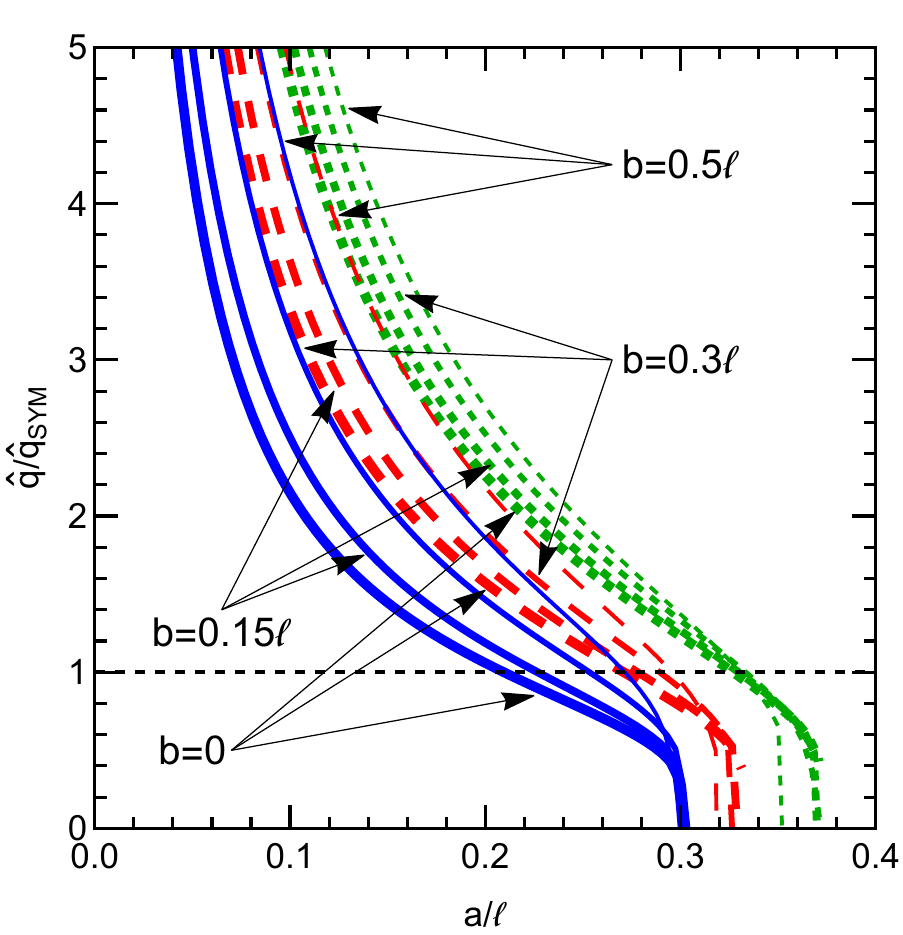}
			\caption{A}
		\end{subfigure}\hfill
		\begin{subfigure}{0.49\linewidth}
			\includegraphics[width=1\linewidth]{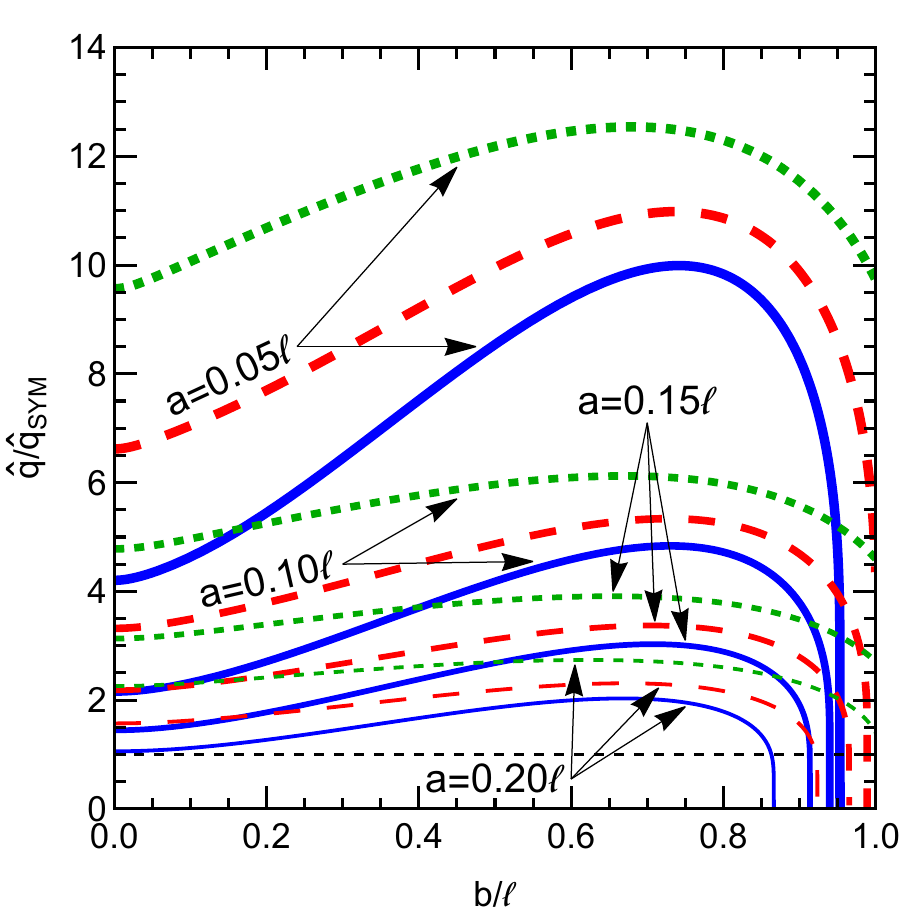}
			\caption{B}
		\end{subfigure}
		\caption{A) The dependence of the ratio $\hat{q}$ in the Kerr-$AdS_5$ to $\hat{q}_{\rm SYM}$ in the planar AdS black hole on the rotational parameter $a$ for different $b$ and $T_{\rm H}$: $T_{\rm H}=0.17$GeV (solid blue), $T_{\rm H}=0.20$GeV (dashed red), $T_{\rm H}=0.30$GeV (dotted green). B)  $\hat{q}/\hat{q}_{\rm SYM}$ as a function of the rotational parameter $b$ for different  $a$ and  $T_{\rm H}$. The rotational parameter $a$ is fixed as $a= 0.20\ell$ for  $T_{\rm H}=0.17, 0.20, 0.30$GeV.}
		\label{fig:qhat-qsym}
	\end{figure}
	It is instructive to see the ratio $\hat{q}/\hat{q}_{\rm SYM}$  in terms of a rotational parameter.  In Figs.~\ref{fig:qhat-qsym} {\bf A,B} the quantity $\hat{q}/\hat{q}_{\rm SYM}$  is depicted as a function of one rotational parameter ($a$ or $b$), while another rotational parameter varies. We  plot this for various values of $T_{\rm H}$. From  Fig.~\ref{fig:qhat-qsym}{\bf A} we see that the jet-quenching parameter  in the Kerr-$AdS_{5}$ black hole is larger than that one in the AdS black hole with a planar horizon. It also can be found from Fig.~\ref{fig:qhat-qsym}{\bf A} that $\hat{q}$ in Kerr-$AdS_{5}$ with a fixed $b$ decreases as the parameter $a$ increases.
	Fig.~\ref{fig:qhat-qsym}{\bf B} shows that the dependence of  $\hat{q}/\hat{q}_{\rm SYM}$ on the rotational parameter $b$  is non-monotonic. For some fixed $a$  the quantity  $\hat{q}/\hat{q}_{\rm SYM}$ increases  as $b$ increases reaching its maximal value and then decreases. This is related to the definition  of the "light-cone" coordinates (\ref{lightcone}),  which yields the emphasis of the parameter $a$.  
	
	\begin{figure}[h!]
		\centering
			\includegraphics[width=0.43\linewidth]{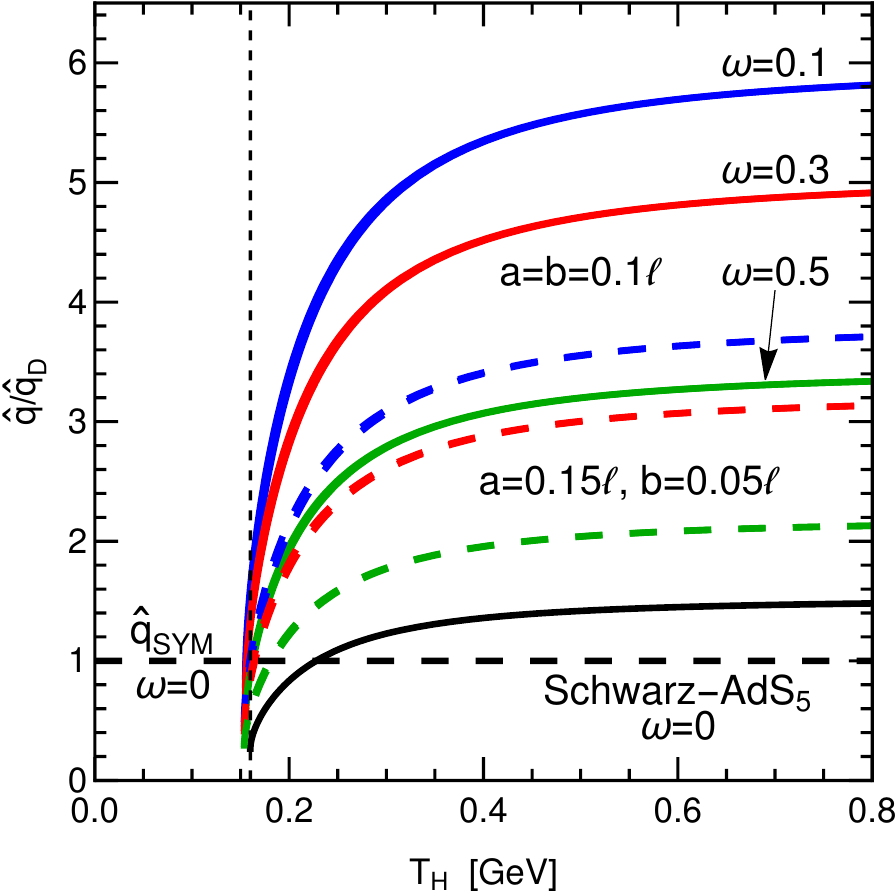}
		\caption{ $\hat{q}/\hat{q}_{D}$ as a function of  $T_{\rm H}$ for various values of the angular velocity of the D-instanton $\omega=0.1, 0.3, 0.5$  and fixed  parameters  $a$ and $b$ of  Kerr-AdS$_5$: $a=b=0.1\ell$ (solid) and $a=0.15\ell$, $b=0.05\ell$ (dashed). The cases of $\hat{q}/\hat{q}_{D}$, where $\hat{q}$ corresponds to the AdS-Schwarzschild and AdS black brane are shown by black solid curve and dashed line, correspondingly ($\omega=0$).}
		\label{fig:qhat-instanton}
	\end{figure}
	
		It is interesting to compare $\hat{q}$ in Kerr-$AdS_{5}$ with the jet-quenching parameter in other holographic backgrounds with the rotation.  For this we focus on the rotating D-instanton background from \cite{Chen:2022obe}. The rotating D-instanton background is characterized by the angular velocity  $\omega$  and the instanton density $q$. In Fig.~\ref{fig:qhat-instanton} we show the ratio $\hat{q}/\hat{q}_{D}$ as a function of $T_{\rm H}$, where $\hat{q}$ is the jet-quenching parameter in the Kerr-$AdS_{5}$ black hole and $\hat{q}_{D}$ is the jet-quenching parameter in the D-instanton  background. We plot $\hat{q}/\hat{q}_{D}$ in terms of $T_{\rm H}$ for different values of the angular velocity $\omega$ and the rotational parameters $a$ and $b$. From Fig.~\ref{fig:qhat-instanton} we  observe that for all $\omega$ and $a,b$ the ratio  $\hat{q}/\hat{q}_{D}$ turns to have a common form: it increases up to some $T_{\rm H}$, and then  takes a constant value. Thus, one can conclude that both jet-quenching parameters $\hat{q}$ and $\hat{q}_{D}$ have the same behaviour at high temperatures.

     We also present  $\hat{q}/\hat{q}_{D}$ in terms of $T_{\rm H}$, where $\hat{q}$ is  taken for the AdS black brane (black dashed line) and the Schwarzschild-$AdS_5$ background (black solid curve), in these cases we set $\omega=0$ for the D-instanton angular velocity. Note that the values of the D-instanton density $q$ are taken in terms of  $\ell$. In fact, we change $q$ in Fig.~\ref{fig:qhat-instanton} from $0$ to $\ell^4$, however the dependence of $\hat{q}/\hat{q}_{D}$  on $q$ is almost negligible. We see that the ratio $\hat{q}/\hat{q}_{D}$ has the similar dependence on $T_{H}$  both for  Kerr-$AdS_{5}$ and  the rotating D-instanton background.

	\section{Conclusions and discussion}
	
	In this paper, we have investigated holographic Wilson loops in the Schwarzschild-$AdS_5$ and Kerr-$AdS_5$ black holes. The both backgrounds have the conformal boundary $R\times S^{3}$ and are holographically dual  to the non-rotating and rotating $\mathcal{N}=4$ SYM quark-gluon plasma, correspondingly. In particular, we have calculated the temporal and light-like Wilson loops in the Schwarzschild-$AdS_5$ and Kerr-$AdS_5$ backgrounds.
	
	From the holographic temporal Wilson loops we have found  the quark-antiquark potentials. We have shown that the expressions for the potentials in both backgrounds contain the linear and Coulomb-like terms.
	At temperatures above the critical one ($T_{\rm H}\geq0.17$ GeV) we have observed the Coulomb-like behaviour (see Figs.~\ref{fig:Vqq-Schwar}{\bf A}, ~\ref{fig:VqqKerr-L}{\bf A}). We have estimated the coefficients of the Coulomb terms  both in the Schwarzschild-$AdS_5$ and Kerr-$AdS_5$ backgrounds from fitting of $V_{q\bar{q}}$. For the non-rotating case ($a=b=0$) we have seen that  the distance between quak-antiquark pair can be decreased either by increasing the temperature $T_{\rm H}$  or reducing the angle $\theta$, see Fig.~\ref{fig:Vqq-Schwar}{\bf A}. The same dependence is inherited for the non-zero rotating parameters,  Fig.~\ref{fig:LKerr-C}.  Note that the similar deformation of the string profile was also observed in \cite{Giataga22} for  rotating mesons in a static background. We have seen that the rotation increases the values of  the quark-antiquark potential $V_{q\bar{q}}$ comparing to the Schwarzschild-$AdS_5$ case at the same interquark distance.  At  high temperatures ($T_{\rm H}=0.30$ GeV) we have observed that $V_{q\bar{q}}$ in the Kerr-$AdS_5$ background becomes closer to $V_{q\bar{q}}$ in the Schwarzschild-$AdS_5$  black hole Fig.~\ref{fig:VqqKerr-L}{\bf A} at least for certain values of the angle $\theta$ ($\Theta$).  
	
	Considering the holographic light-like Wilson loops, we have calculated the jet-quenching parameters $\hat{q}$.  For the Schwarzschild-$AdS_5$ black hole we have found that the analytic expression for $\hat{q}$ at high temperatures and  $\theta \sim 0$ has a cubic dependence on $T_{\rm H}$, which is  similar to that one in the  AdS  black brane (with a planar horizon) \cite{LRW}. We have also observed this from the dependence of $\hat{q}/T^{3}_{H}$ on $T_{\rm H}/T_{\rm H}^{min}(0,0)$ (see Fig.~\ref{fig:qhat-T-Schwarz} {\bf B}). Like we have seen for $V_{q\bar{q}}$, the value of the jet-quenching parameter $\hat{q}$ also depends on $\theta$. For example, one can obtain a value of $\hat{q}$ at  $\theta < \pi/9$, which is smaller than in the AdS black brane.
	
	In the case of Kerr-$AdS_{5}$ we have found that the jet-quenching parameter increases with the rotation, see~Fig.\ref{fig:qhat-T-Kerr}{\bf A}. However, at high temperatures we still have the dependence $\hat{q} \sim  \kappa_{\rm rot} T^{3}_{\rm H}$ with  $\kappa_{\rm rot}$ defined by values of $a$ and $b$. Thus, the cubic dependence of $\hat{q}$ on  $T_{\rm H}$ takes place at high temperature for the rotating and non-rotating cases. 
	In both cases, we have observed a strong dependence on the angle $\theta$ (or $\Theta$), which is related to the geometries. Remarkably, in \cite{Chen:2022obe} it was found the jet-quenching parameter in the rotating case (the rotating D-instanton background) takes also larger values.
	 
An interesting future direction could be the generalisation to a charged Kerr-$AdS_{5}$ background \cite{Cvetic}, which corresponds to the case with a non-zero the chemical potential. Another interesting problem would be a study of spacial Wilson loops in the Kerr-$AdS_{5}$  and  the Kerr-Newmann-$AdS_{5}$ black holes and comparing with the lattice results for the rotating quark-gluon plasma \cite{Braguta:2020,Braguta:2021}.  It would be useful to consider holographic probes moving along a circle in the rotating QGP as it was discussed for non-rotating black branes in \cite{BitaghsirFadafan:2008adl,Atashi:2019mlo}.

	\section*{Acknowledgments}
	We are grateful to Irina Aref'eva, Eric Gourgoulhon, Yury Ivanov, Timofey Rusalev and  Oleg Teryaev for the valuable discussions.  We thank Olesya Geytota for the participation at early stages of the project. We also thank Dimitrios Giataganas and Kazem Bitaghsir Fadafan for the correspondence and useful comments.
	 The present paper were written with the support of the Russian Science Foundation grant No 22-72-10122.

\end{document}